\documentclass[aps,pre,superscriptaddress,showpacs,amsmath,amssymb,amsfonts,altaffillsymbol,twocolumn,nofootinbib]{revtex4-2}

\usepackage{graphicx}
\usepackage{float}
\usepackage{color}
\usepackage{dcolumn}

\graphicspath{{./figures/}}
\definecolor{DarkPurple}{RGB}{130,0,255}
\usepackage{ulem}
\usepackage{tensor}
\usepackage{dsfont}
\usepackage{lipsum} 
\usepackage{multirow}
 
\usepackage{xcolor,colortbl}

\usepackage{bm}

\newcommand{\Up}{\mathbf{U}}
\newcommand{\Dn}{\mathbf{D}}

\begin{document}
\title{
Cooperative effects driving the multi-periodic dynamics of cyclically sheared amorphous solids
}

\author{Asaf Szulc}
\affiliation{Department of Physics, Ben Gurion University of the Negev, Beer Sheva 84105, Israel}

\author{Muhittin Mungan}
\affiliation{Institut f{\"u}r angewandte Mathematik, Universit{\"a}t Bonn, Endenicher Allee 60, 53115 Bonn, Germany}

\author{Ido Regev}
\email[E-mail: ]{regevid@bgu.ac.il}
\affiliation{Department of Solar Energy and Environmental Physics, Jacob Blaustein Institutes for Desert Research, Ben-Gurion University of the Negev, Sede Boqer Campus 84990, Israel}

\date{\today}

\begin{abstract}
Plasticity in amorphous materials, such as glasses, colloids, or granular materials, is mediated by local rearrangements called ``soft spots''. Experiments and simulations have shown that soft spots are two-state entities interacting via quadrupolar displacement fields generated when they switch states. When the system is subjected to cyclic strain driving, the soft spots can return to their original state after one or more forcing cycles. In this case, the system has periodic dynamics and will always repeat the same microscopic states. Here we focus on multi-periodic dynamics, i.e. dynamics that has periodicity larger than the periodicity of the drive, and use a graph-theoretical approach to analyze the dynamics obtained from numerical simulations. In this approach, mechanically stable configurations that transform purely elastically into each other over a range of applied strains, are represented by vertices, and plastic events leading from one stable configuration to the other, are represented by directed edges. An algorithm based on the graph topology and the displacement fields of the soft spots reveals that multi-periodic behavior results from the states of some soft spots repeating after more than one period and provides information regarding the mechanisms that allow for such dynamics. To better understand the physical mechanisms behind multi-periodicity, we use a model of interacting hysterons. Each hysteron is a simplified two-state element representing hysteretic soft-spot dynamics. We identify several mechanisms for multi-periodicity in this model, some involving direct interactions between multi-periodic hysterons and another resulting from cooperative dynamics involving several hysterons. These cooperative events are naturally more common when more hysterons are present, thus explaining why multi-periodicity is more prevalent at large drive amplitudes.
\end{abstract}

\maketitle

\section{Introduction}
Understanding the response of amorphous materials to time-varying mechanical loads is one of the main topics of interest in different fields, such as mechanical engineering, chemical engineering, materials science, geophysics and condensed matter physics. 
Considering the mechanical response of materials, traditionally one distinguishes two types of deformations: elastic and plastic. 
When a material is deformed elastically, the deformation is reversible in the sense that if the applied forces are removed, the material returns
to its initial state. This type of deformation is typically encountered when the applied forces are small. For large forces, materials typically deform plastically, which means that once the forces are released these materials do not return to their original shape and are thus deformed in an irreversible manner.
In recent years it was discovered that under athermal quasi-static (AQS) conditions, plastic deformation can also be reversed ~\cite{Regev2013, Keim2013, Keim2014, Regev2015, Fiocco2013} and hence this present an intermediate type of deformation -- {\it reversible plasticity}. One way in which reversible plasticity
arises, is as a result of oscillatory shear. When oscillatory shear of moderate amplitude is applied to an amorphous material, it experiences transient and irreversible
dynamics, until it responds periodically to the cyclic forcing so that 
the same microscopic configurations reappear periodically, and transitions between configurations involve both elastic and plastic deformations. Thus during cyclic response, the plastic events taking place are repeatable. 

Measured in units of the number of driving cycles, one denotes the length of the transient leading to cyclic behavior as $\tau$, while the period of the response is $T$. In the case of $T>1$, this type of periodic response is called {\it multi-periodic}\footnote{In the case of driven spin-glasses this type of response has also been called {\it subharmonic} \cite{Deutsch2003}.}. It is known that as the amplitude of an applied oscillatory shear approaches a critical amplitude, the transients get increasingly longer, while multi-periodic cycles start to appear. Once the critical amplitude is reached, a transition to diffusive behavior occurs, as a result of which a periodic response is no longer attainable. This transition has been called the {\it irreversibility transition} and it is believed that the strain amplitude at which it occurs coincides with the strain at which the material yields under uniformly increasing shear.
Understanding the mechanism leading to multi-periodic response is of interest, as it emerges at strain amplitudes close to the irreversibility transition~\cite{Regev2013, Regev2015}, and hence may shed light on it. At the same time, a $T>1$ response is also of interest due to the possibility of encoding mechanical memory~\cite{Keim2021, Lindeman2021, Hecke2021, bense2021}.

In amorphous materials, such as glasses, colloids or granular materials, in which the basic constituents are not organized in periodic structures and in which long-range order is absent, plasticity is mediated by local rearrangement events. The regions where these arrangements occur are called ``soft spots'' or ``shear transformation zones''~\cite{Falk1998, Falk1999, Maloney2006, Schall2007, Manning2011, Regev2013, Keim2013, Keim2014}. An isolated soft spot has hysteresis properties: when sheared in the positive direction its constituent particles undergo a localized rearrangement at a strain $\gamma^+$, and when subsequently strained in the opposite direction, this rearrangements is reverted at some strain $\gamma^-<\gamma^+$~\cite{Mungan2019}. When an amorphous material is subject to cyclic deformation, several different soft spots become active, undergoing localized rearrangements. Due to the long-range elastic deformations accompanying these state changes, soft spots interact with each other: the state change of one soft spot can change the switching behavior of other soft spots so that the sequence of these rearrangements determines the mechanical response and hysteresis of the material. 

The goal of this work is to understand the origin of multi-periodic response in cyclically driven amorphous materials. Contrary to previous work that focused on the emergence of multi-periodicity and long transients due to the complexity of the energy landscape \cite{szulc2020}, 
here we focus on the emergence of this response  as a result of the interactions between individual soft-spots.
We achieve this by using a graph-theoretical approach, recently introduced by two of us~\cite{Mungan2019}, which represents the AQS dynamics of driven disordered systems as a directed transition graph, the {\it $t$-graph}~\cite{Mungan2019b,Mungan2019}. The $t$-graph provides a faithful representation of the dynamics that moreover can be extracted from numerical simulations, as described in \cite{Mungan2019, Regev2021}. Its vertices, the {\it mesostates}, are collections of mechanically stable configurations that transform purely elastically into each other over a range of applied strains, while its directed edges are the purely plastic events under which a configuration belonging to one mesostate transits to another one.  

The $t$-graph allows us to trace the response to arbitrary driving protocols, particularly the periodic response to cyclic driving. Combining this with spatial information about soft spots that change states during the corresponding mesostate transitions permits us to identify the active soft spots during periodic response. We focus first on $T=2$ multi-periodic cycles that we extracted from the $t$-graphs. 
We then use a specialized algorithm to identify and tag the active soft spots and investigate their state switching behavior as the cycle is traversed. We observe that some of the active soft spots switch between their two states during each forcing cycle and thus have period $1$, while others have period $2$, switching their states in one forcing cycle and reverting back in the other. The latter are what sets the period of the cycle to be $T=2$. 
We find that the most common mechanism that causes the period-$2$ switching behavior of some soft spots is that the strain thresholds $\gamma^+,\gamma^-$, of these soft spots periodically fluctuate between values that are larger (smaller) than the maximal (minimal) strain: when the strain thresholds are too large, the soft spots cannot change states during the driving cycle, while when their values fall within the range of the driving they can switch states and hence are {\it active}. We will call this phenomenon {\it threshold oscillations} and they emerge as a result of interactions with other soft spots that are active during a forcing cycle.

While the $t$-graph approach combined with our ability to identify soft spots changing states during a transition, allows us to determine how the switching fields of a given soft spot change as a result of the state changes of some other soft spots, this turns out not to provide sufficient information to determine how the state change of any one soft spot changes the strain thresholds of {\it all} other active soft spots. We thus obtain only a limited amount of information about the pair-wise interaction between active soft spots present. To remedy this, we turn next to a model of interacting hysteresis elements, the {\it interacting Preisach model}, which was originally introduced by Hovorka and Friedman \cite{Hovorka2005}, and subsequently revisited within the context of memory formation in driven mechanical systems \cite{Lindeman2021, Keim2021, bense2021, Hecke2021}. 

In this model, a soft spot corresponds to a {\it hysteron}, a basic hysteresis unit with maximal and minimal strain thresholds. 
We show that a set of hysterons with strain thresholds and interactions between them chosen at random, reproduces dynamics that is qualitatively similar to the dynamics observed in particle simulations, exhibiting $T = 2$ multi-periodic response where some hysterons switch with period $2$ while others switch with period $1$. 
More importantly, in the interacting Preisach model, we can follow the effects of interactions between hysterons at all times which allows us to uncover the mechanisms that lead to multi-periodic response. 

We then study multi-periodic response in a system of $100$ interacting hysterons by artificially "knocking-out" the hysterons, {\it i.e.} deactivating them by disabling their switching behavior, and then allowing individual hysterons to activate and join the dynamics. We show that after several hysterons are activated, the addition of one more hysteron leads to a transition from a $T=1$ to a $T=2$ cycle. 
Upon examination, we find that period-2 oscillations of individual hysterons arises due to two mechanism involving fluctuations of the strain thresholds. The first mechanism, involves a temporary \textit{freezing} of the hysteron during part of the cycle, which occurs when interactions shift the strain threshold of the hysteron to the extent that $\gamma^+$ becomes smaller than $\gamma^-$. The second mechanism, that can occur concurrently to the first, involves a periodic fluctuation of the strain threshold of a hysteron around the extreme values of the driving force that in the following will be referred to as {\it strain threshold oscillations}. In a small system, the switching of one hysteron can cause the strain threshold of another hysteron to temporarily become larger than the forcing amplitude, which causes these oscillations. 
In a complex, many-hysteron, system such behavior can also occur due to a cooperative effect involving several hysterons. This mechanism involves two steps: the first step is a {\it cascade} of strain threshold shifts that causes some hysterons to switch prematurely, which then causes the $\gamma^+$ ($\gamma^-$) of one hysteron to go above (below) the strain amplitude. The second step is a {\it scrambling}~\cite{Hecke2021, bense2021} of the switching order of the hysterons involved in the cascade between the different forcing cycles. The periodic scrambling then causes the strain threshold of this hysteron to oscillate around the forcing amplitude, making the switching pattern of the hysteron multi-periodic. In the following, we will call this effect {\it cascade-scrambling}.

While there are indications that these mechanisms may also be present in the particle simulations, we can currently only confirm the existence of strain threshold oscillations. Finding the other mechanisms in particle simulations will be left for future work.
However, the relation between strain threshold oscillations and the interaction between several soft spots, provides an explanation for the fact that multi-periodic cycles in amorphous solid at strain amplitudes that are close to the irreversibility transition, where the number of active soft spots becomes large. 

The paper is organized as follows: In section \ref{section_2}, we first analyze cycles obtained from a particle simulation (details on the particle simulation can be found in Appendix~\ref{App0}). We show how $t$-graphs are used to locate and identify the different soft spots active during a $T=1$ cycle. We then explain how this is extended to the more complicated $T>1$ cycles, and use the strain thresholds obtained from this analysis to show that $T>1$ is a result of oscillations of the strain thresholds of some of the soft spots around the forcing amplitude. In section \ref{section_3} we use a model of interacting hysterons as means for understanding the origin of these oscillations. In section \ref{section_4}, we show how interactions cause such oscillations in a simple configuration of three hysterons with parameters that were specifically chosen to allow multi-periodicity. In section \ref{section_5} we examine a system of $100$ hysterons with random strain thresholds. By deactivating the hysterons and then allowing hysterons to become active one by one, we show that the system starts at $T=1$ and transitions to $T=2$ when a specific hysteron is added. In the example shown, the set of hysterons active in both $T=1$ and $T=2$ cycles is practically the same, and the difference in the dynamics comes from a shift in the strain thresholds of all the hysterons due to the newly activated one. The induced perturbation reveals two new mechanisms that contribute to the emergence of multi-periodic cycles in large systems.
In section \ref{sec:comparePreisachParticle} we compare these results with the observations in particle simulations and discuss indications for the existence of these mechanisms in these simulations. We summarize and discuss open questions and future directions in section \ref{sec:discussion}. 

\section{Analysis of multi-periodic cycles using $t$-graphs}
\label{section_2}
\begin{figure*}[tb]
\begin{center}
\includegraphics[width=0.90\linewidth]{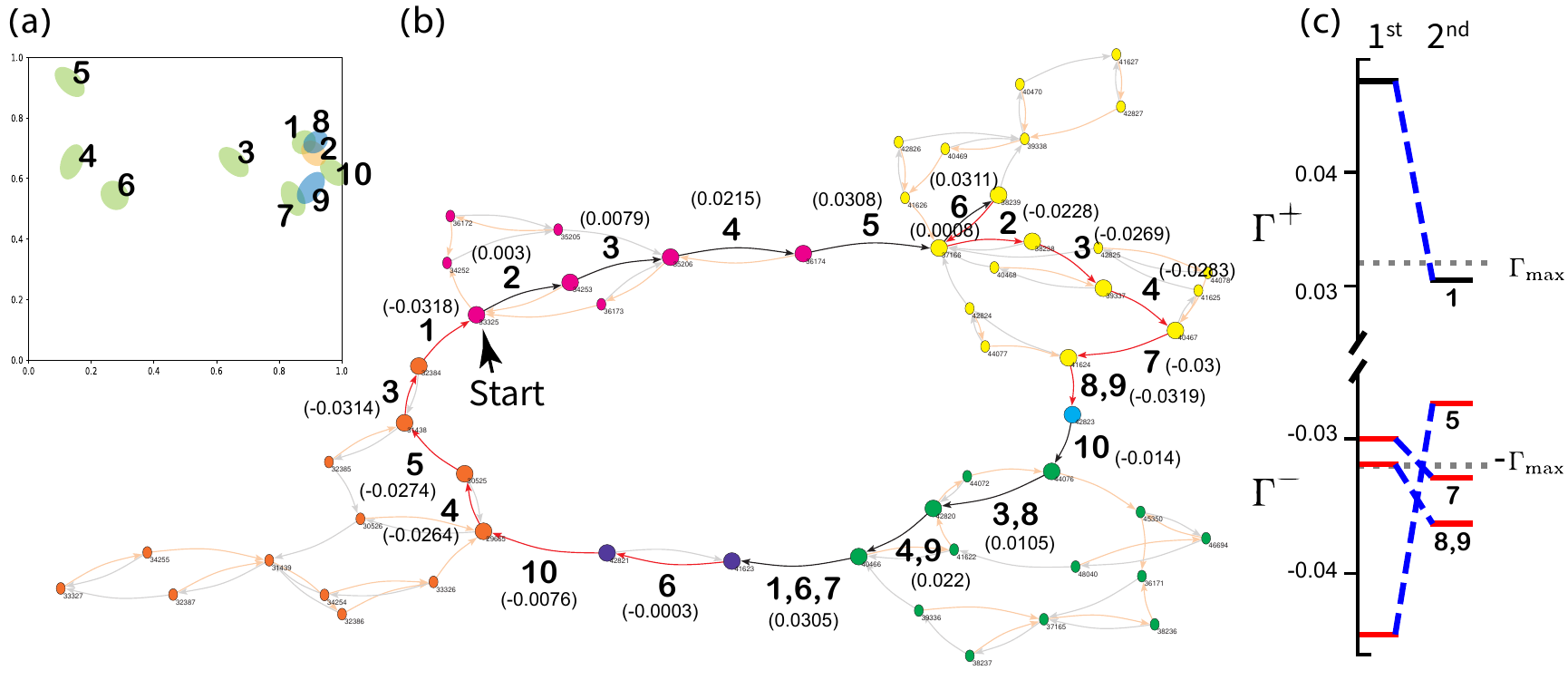}
\caption{(a) The spatial positions of the soft spots activated during a $T=2$ cycle generated by a particle simulation. The ellipses enclose the four center particles of each soft spot, where the different colors aim to distinguish between overlapping ellipses. (b) A graph representation of the cycle. Black and gray arrows mark the transitions under increasing strain, while transitions as a result of decreasing strain are shown as red and orange arrows.
The transitions constituting the cycle are colored black and red and the numbers next to them indicate the soft spots switching in the transition. The strain initiating the transition is written next to it in parentheses. An arrow marks the beginning of the first  driving cycle. (c) The difference in $\Gamma^+_i$ (black) and $\Gamma^-_i$ (red) values of several period-$2$ soft spots in the two driving cycles.
}
\label{Fig1}
\end{center}
\end{figure*}
The periodic response of an amorphous solid subject to oscillatory shear at relatively low strain amplitudes, is composed of soft spots switching back and forth between two states~\cite{Mungan2019}. 
For a $T=1$ cycle in which the periodic response has the same periodicity as the driving cycle, each soft spot switches twice during the cycle such that when the driving completes a cycle, the system returns to its initial configuration. In the following, we will represent the two states of the $i$th soft spot as $s_i = 0,1$. A soft spot can either be in state $s_i=0$ or $s_i=1$, and the labeling will be such that it switches from $0$ to $1$ during increasing driving, and from $1$ to $0$ when the driving is decreasing. We denote the strain at which soft spot $i$ switches from $s_i=0$ to $s_i=1$ due to an increase in the strain, by $\Gamma^+_i$, and the strain at which the same soft spot transitions back from $s_i=1$ to $s_i=0$ due to a decrease in the strain, by $\Gamma^-_i$. Since we are considering AQS dynamics, so that the response is rate-independent, it suffices to assume that the driving has a triangular  wave form, oscillating in one period as $-\Gamma_{\max} \to +\Gamma_{\max} \to -\Gamma_{\max}$, where $\Gamma_{\max}$ is the driving amplitude.
Each driving cycle thus begins with a segment of increasing strain, followed by a segment of decreasing strain. We will denote these two segments as $\nearrow$ and $\searrow$, respectively. 

Fig.~\ref{Fig1}(a) shows the locations of the $10$ soft spots that are active in an example of a $T=2$ cycle. Table~\ref{tab:Fig3_summary} depicts the \textit{switching-patterns of  active soft spots, i.e.} the state $s_i$ of each soft spot $i$ at the ends of the  different segments of the cycle. For future reference, we will denote by  
${\bf S} = ( s_1, s_2, ..., s_{10})$ the soft spot configuration attained at  
the end of the segments $\nearrow$ and $\searrow$ of the driving cycle.
We can see that there are three period-$1$ soft spots -- \textit{3}, \textit{4}, and \textit{6} -- that switch in all four segments.
If these were the only soft spots active during the cycle, the configuration would have returned to its original state after the first driving cycle.
 However, besides the period-$1$ soft spots, there are also seven period-$2$ soft spots -- \textit{1}, \textit{2}, \textit{5}, \textit{7}, \textit{8}, \textit{9}, and \textit{10} -- that change their state only during one of the driving cycles. Another feature that can be observed in Fig~\ref{Fig1}(a), is that some soft spots overlap. Possible implications of this feature will be discussed later-on.

\subsection{Using the $t$-graph to infer soft spot locations and their interactions}

To understand the origin of the period-$2$ soft spot switching mechanism, we need to understand how the switching of one or more soft spots affects the switching behavior of other soft spots, such that they can switch only during part of the cycle. This requires identifying the strain thresholds of a given soft spot, as {\it the background}, i.e. the states of the other soft spots, change. 
To identify these thresholds, we represent the plastic transitions encountered by the material during the periodic response using a $t$-graph. 
In terms of the sheared amorphous solid, the nodes, or mesostates, of the $t$-graph are collections of mechanically stable particle configurations that deform purely elastically into each other as the applied shear strain $\Gamma$ is changed. Thus a mesostate $A$ deforms elastically over a range of strains $\Gamma$ that we denote by $\Gamma^-[A] < \Gamma < \Gamma^+[A]$. Here $\Gamma^\pm[A]$, are the strains at which a plastic event occurs. When $\Gamma = \Gamma^+[A]$, a plastic event that is triggered by one soft spot changing its state from 
$s_i= 0 \to 1$, leads to a new configuration that is stable at $\Gamma = \Gamma^+[A]$. This new configuration must belong to some other mesostate $B$. A similar situation arises when $\Gamma = \Gamma^-[A]$, where a plastic event triggered by the state change of a soft spot
$s_i= 1 \to 0$ leads to a new mesostate $C$. The process is repeated for mesostates $B$ and $C$, such that the mesostates and the transitions between them form the $t$-graph. Details on the extraction of $t$-graphs from simulations of sheared amorphous solids can be found in \cite{Mungan2019} and are summarized in Appendix \ref{AppA}.
The $t$-graph together  with the stability ranges $\Gamma^\pm[A]$ of its mesostates,  is a map-like representation of  the AQS response under arbitrary shear protocols\footnote{Since we are only considering one-parameter deformations, the applied shear has to be along one shearing line in two-dimensions or along a shearing plane in three dimensions.}~\cite{Mungan2019}. Specifically, graph-cycles correspond to periodic behavior, and the transient dynamics is represented as a path on the graph leading to a graph-cycle. Fig.~\ref{Fig1}(b) shows the $t$-graph representation the $T=2$ cycle whose soft spots are shown in Fig~\ref{Fig1}(a). Here arrows colored in black/gray (red/orange) represent transitions when the strain is increased (decreased). 

As noted already, in sheared amorphous solids the plastic events connecting two mesostates correspond to soft spots in the sample that undergo a local rearrangement. Thus by combining the $t$-graph with spatial information of the particle displacements during a plastic event, one can, in principle, identify the soft spots that change during a graph transition. In particular, if the transition involves the state change of a single soft spot, we can  also attribute the corresponding thresholds strain $\Gamma^\pm[A]$ of the mesostate at which this event occurs to the particular soft spot changing state. Fig.~\ref{Fig1}(b) is an excerpt of a $t$-graph obtained from simulations of a sheared amorphous solid and shows the multi-periodic response with $T=2$. The transitions making up the cycles are highlighted in black and red, corresponding to the plastic events under increasing and decreasing strain and the numbers next to a transition identify the corresponding soft spots that changed their state. 

While constructing the $t$-graph involves a simple, easy to define, algorithm, identifying the different soft spots active during the transition and their locations, can be challenging since in many cases, plastic events include more than one soft spot. In~\cite{Mungan2019}  soft spots were identified by comparing their quadroplar displacement fields~\cite{Maloney2006} by eye and
using the fact that the cycle included many sub-cycles in which the same soft spots were active, so that a soft spot exists independently of a particular plastic event. For this work, we have created an algorithm that automates this process and identifies all the soft spots in a cycle which has perfect or near-perfect loop return-point memory ($\ell$RPM) property, to be discussed below. 
 
\subsubsection{$T\,=\,1$ Cycles}
 
For periodic cycles that have the $\ell$RPM property, there is a well-defined hierarchy of sub-cycles \cite{Mungan2019b,Mungan2019}. 
The main cycle is composed out of a lower and an upper mesostate, the endpoints of the cycle, which we will denote by ${\bf S}_{\rm zero}$ and ${\bf S}_{\rm one}$. Starting in the lower endpoint ${\bf S}_{\rm zero}$, the upper endpoint is reached by increasing the strain to some value $\Gamma_{\max}$, while a  subsequent decrease of strain to some value $\Gamma_{\min}$ brings the system back to ${\bf S}_{\rm zero}$. Thus under repeated applications of $ \Gamma_{\min} \to \Gamma_{\max} \to \Gamma_{\min}$, the graph-cycle with endpoints ${\bf S}_{\rm zero}$ and ${\bf S}_{\rm one}$ is traversed. The $\ell$RPM property assures that if starting again from ${\bf S}_{\rm zero}$, but increasing the 
strain this time up to some value $\Gamma < \Gamma_{\max}$, so that some state ${\bf S}$ is reached, then a sufficiently large decrease of the strain will always bring the system back to ${\bf S}_{\rm zero}$. Thus ${\bf S}_{\rm zero}$ and ${\bf S}$ are the endpoints of a sub-cycle\footnote{Similarly starting from the upper endpoint ${\bf S}_{\rm one}$ and decreasing the strain to some value 
$\Gamma > \Gamma_{\min}$, a subsequent increase of strain will eventually return the system to ${\bf S}_{\rm one}$.}.

In terms of the soft spots, one can associate with a cycle a set of active soft spots that change their states as the cycle is traversed, so that at the lower endpoint ${\bf S}_{\rm zero}$ of the cycle all active soft spots are in their $0$ state, ${\bf S}_{\rm zero} = ( 0, 0, \ldots, 0 )$, while at the upper endpoint they are in their $1$ state, ${\bf S}_{\rm one} = (1, 1, \ldots, 1)$. Thus when $\ell$RPM is present, in the corresponding sub-cycles only a subset of theses soft spots are active. 
Consequently, a given soft spot will be active in multiple sub-cycles and its switching behavior  will depend on the states of the other active soft spots. Thus by using the $t$-graph to decompose a cycle into its sub-cycle, combined with the displacement fields of the plastic events, we can identify the soft spots that are active in each of the transitions. 
In some cases, this also allows us to determine how the switching strains of active soft spots depend on the states of the other soft spots. We have developed a software tool {\tt identSoft} that identifies soft spots in cycles with perfect or near-perfect $\ell$RPM, and infers soft spots interactions. The details are provided in Appendix~\ref{AppB}. 

\subsubsection{$T > 1$ Cycles}
%
\begin{table}[tb]
\begin{center}
\bigskip
\begin{tabular}{c*{10}{c}r}
\hline
Soft spot         & \cellcolor{yellow!50}1 & \cellcolor{yellow!50}2 & 3 & 4 & \cellcolor{yellow!50}5 & 6 & \cellcolor{yellow!50}7 & \cellcolor{yellow!50}8 & \cellcolor{yellow!50}9 & \cellcolor{yellow!50}10 \\
\hline
$\nearrow$   & \cellcolor{yellow!50}0 & \cellcolor{yellow!50}1 & 1 & 1 & \cellcolor{yellow!50}1 & 1 & \cellcolor{yellow!50}1 & \cellcolor{yellow!50}1 & \cellcolor{yellow!50}1 & \cellcolor{yellow!50}0 \\
$\searrow$ & \cellcolor{yellow!50}0 & \cellcolor{yellow!50}0 & 0 & 0 & \cellcolor{yellow!50}1 & 0 & \cellcolor{yellow!50}0 & \cellcolor{yellow!50}0 & \cellcolor{yellow!50}0 & \cellcolor{yellow!50}0  \\
$\nearrow$   & \cellcolor{yellow!50}1 & \cellcolor{yellow!50}0 & 1 & 1 & \cellcolor{yellow!50}1 & 1 & \cellcolor{yellow!50}1 & \cellcolor{yellow!50}1 & \cellcolor{yellow!50}1 & \cellcolor{yellow!50}1  \\
$\searrow$ & \cellcolor{yellow!50}0 & \cellcolor{yellow!50}0 & 0 & 0 & \cellcolor{yellow!50}0 & 0 & \cellcolor{yellow!50}1 & \cellcolor{yellow!50}1 & \cellcolor{yellow!50}1 & \cellcolor{yellow!50}0  \\
\hline
\end{tabular}
\end{center}
\caption{\label{tab:Fig3_summary} Switching patterns of the $10$ soft spots active in the $T = 2$ multi-periodic cycle shown in Fig.~\ref{Fig1}. 
The symbols $\nearrow$ and $\searrow$ represent the consecutive four segments of increasing and decreasing strain of the two driving periods. 
Soft spots with a period-$2$ switching pattern are marked in yellow.}
\end{table}
Multi-periodic cycles are more complex since they do not have the $\ell$RPM property and thus, using the topology is less straightforward. However, the mesostates belonging to such a cycle can be separated into clusters of $\ell$RPM-like hierarchically organized cycles and sub-cycles. 
These clusters are sets of mesostates that are all part of one periodic cycle that obeys a near-perfect RPM, meaning that transitions out of the cycle or its sub-cycles first are from the endpoints of the cycle only.  
Such a decomposition is shown in Fig.~\ref{Fig1} where mesostates belonging to the same clusters are drawn as vertices of the same color. 
For each of the clusters we can use {\tt identSoft} to find the soft spots that are active.
The identification of soft spots involved in transitions between clusters has to be handled separately. In this case, special
care has to be given to transitions  involving more than one soft spot, {\it avalanches}, since  
identifying all the individual soft spots of the avalanche may requires searching for ``nearby'' loops, {\it i.e.} graph-loops that can be reached by a few transitions from the cycle of interest, and in which they are active individually, that is, they are not part of an avalanche. Details are provided in Appendix~\ref{AppB}.

The $t$-graph shown in Fig.~\ref{Fig1}, was extracted from a large graph generated by particle simulations. It corresponds to a  $T=2$ cycle with the lowest number of mesostates. The label "Start" marks the beginning of the driving cycle.  Fig.~\ref{Fig1}(c) shows the difference in $\Gamma^+$ (black) and $\Gamma^-$ (red) values of selected soft spots in the first and second driving cycles. We can see from Fig.~\ref{Fig1}(c) that for most of the soft spots that have period $2$, {\it e.g.} soft spots \textit{1},\textit{5},\textit{7},\textit{8} and \textit{9}, the strain thresholds $\Gamma^\pm_i$ oscillate around $\pm\Gamma_{\max}$. More specifically, for soft spot \textit{1}, $\Gamma^+_1 > \Gamma_{\max}$ in the first driving cycle but $\Gamma^+_1 < \Gamma_{\max}$ in the second driving cycle. Consequently, soft spot $\textit{1}$ switches only in the second driving cycle. Similarly, in the case of soft spots \textit{5}, \textit{7}, \textit{8}, and \textit{9}, their lower strain threshold $\Gamma^-_i$, oscillates around  $-\Gamma_{\max}$ between the first and the second driving cycles. We will refer to such processes as \textit{threshold oscillations}. 
 
The dynamics of soft spots \textit{2} and \textit{10} are somewhat different. The measured values of $|\Gamma^+_{10}|$ and $\Gamma^+_{2}$ are much smaller than $\Gamma_{\max}$, that indicates that it is not likely that they perform oscillations around $\Gamma_{\max}$. Furthermore, even if we increase the strain to values much larger than $\Gamma_{\max}$, we cannot find an activation threshold for these soft spots, in the second forcing cycle. This indicates that both \textit{2} and \textit{10} are effectively not present in the second forcing cycle. On the other hand, we observe that the measured value of $\Gamma^+_{10}$ is significantly lower than the measured value of $\Gamma^-_{10}$. This indicates that these soft spots may be frozen due to a mechanism which causes $\gamma_+$ to becomes equal to $\gamma_-$ in the sense discussed in the introduction.
How such a mechanism can arise, will be discussed in detail below. 

\section{Preisach model with random interactions}
\label{section_3}
The Preisach model is the simplest model of hysteresis. It was originally proposed to model hysteresis in magnetic systems, but has been recently also used to study cyclic response in periodically driven amorphous materials and other mechanical systems~\cite{Mungan2019a, Paulsen2019, Keim2020, bense2021}. In this model, the basic unit of
hysteresis is a two-state system, the {\it hysteron} that when subject to a time varying driving switches between two states, "0" and "1", in a hysteretical manner. The pair of switching fields $\gamma^+_i$ and $\gamma^-_i$ at which the transitions of a hysteron $i$ from $0 \to 1$, respectively $1 \to 0$ occur, characterize the hysteron. Hysteretic behavior occurs when $\gamma^-_i < \gamma^+_i$, so that over this range of values of the driving the hysteron can be  in state $0$ or $1$, and the actual value depends on its driving history.

In the original formulation of the Preisach  model, the hysterons do not interact with each-other and thus the dynamics of a collection of hysterons, each independently subjected to the same periodically changing driving, will result in a cyclic response whose period is $T=1$.
However, as is the case for spin-glass models~\cite{Deutsch2003}, when interactions between hysterons are introduced, so that the state of one hysteron $i$ may facilitate or hinder the switching behavior of another hysteron $j$ by means of changing the values of its  strain thresholds, multi-periodic cycles, {\it i.e.} cycles with $T > 1$ start to appear, as was shown  recently~\cite{Keim2021, Lindeman2021}. In the following, we will refer to models of interacting hysterons as \textit{iPreisach} models.

\subsection{From soft spots to hysterons}
\begin{figure}[tb]
\begin{center}
\includegraphics[width=0.90\linewidth]{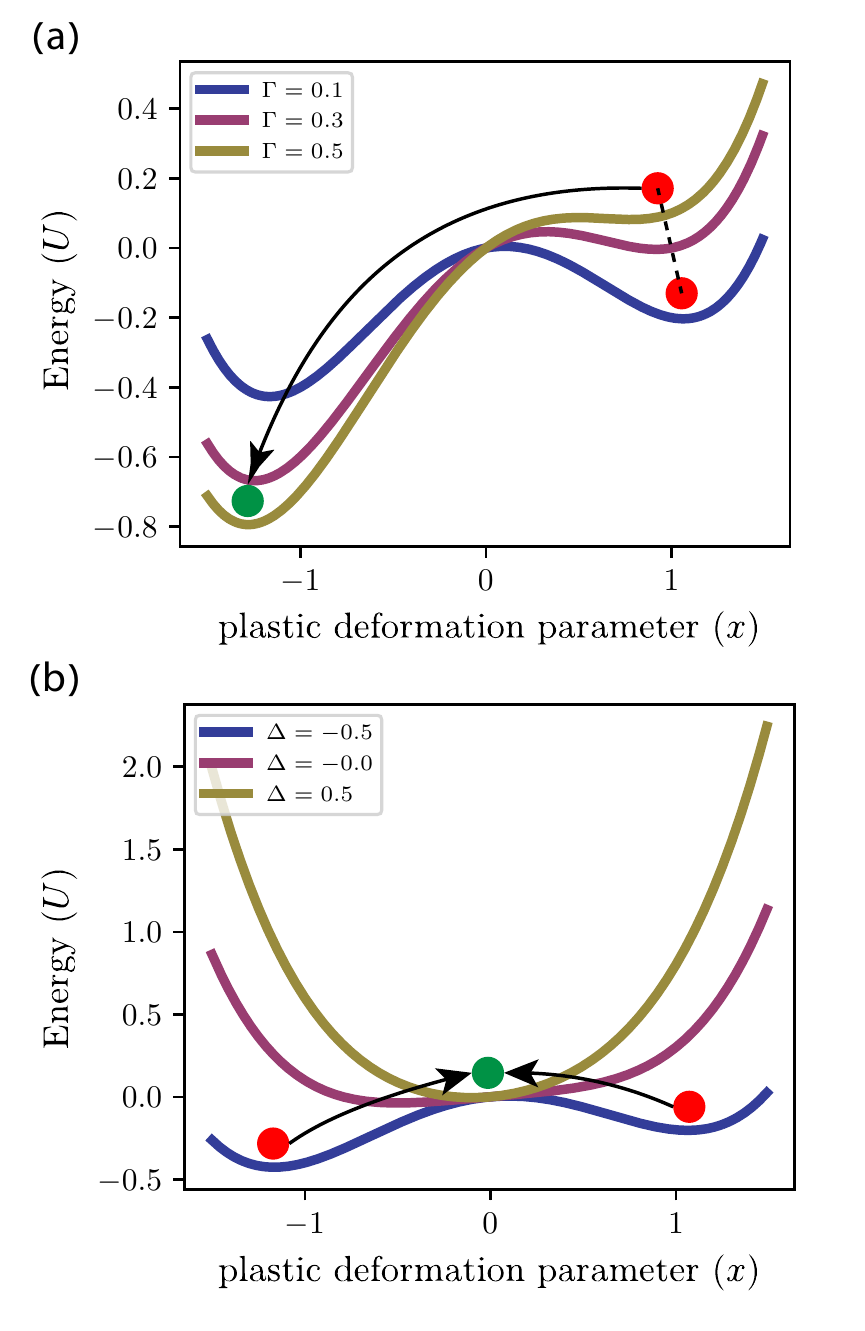}
\caption{A typical asymmetric double well potential $U = \Lambda x^4 + \Delta x^2 +\Gamma x$ ($\Lambda=0.2$), showing (a) how a hysteron state switches, and (b) two minima of the hysteron annihilating and creating a new stable minimum. The initial state is marked with a red circle and the final state is a green circle.
}  
\label{Fig2}
\end{center}
\end{figure}

The iPreisach model can be thought of as an abstract representation of a set of active soft spots in an amorphous solid that interact with each other.
As we saw in the previous section, each soft spot can be regarded as a two-level system, and its potential energy thus can be approximated by a double-well potential, {\it i.e.} a soft spin, as considered by Puglisi and Truskinovsky \cite{Puglisi2002}. 
Here the two minima of the potential correspond to the states $s_i=0,1$ of the $i$th soft spot.
In Fig~\ref{Fig2} we see an example of a double-well potential of the form
\begin{equation}
U(x) =  \Lambda x^4 + \Delta x^2 + \Gamma x, 
\end{equation}
as well as the effect of changing the external driving $\Gamma$. Here $x$ is a coordinate representing the state of the soft spot, $\Lambda$ and $\Delta$ are constants. 
Starting from the left minimum and increasing $\Gamma$, the potential is distorted and when it reaches a critical value $\Gamma^+$, the minimum becomes unstable (a saddle-node bifurcation) and the system transitions to the other minimum. Similarly, starting from the other minimum and decreasing $\Gamma$, at a critical value $\Gamma^-$, the minimum is switched again [Fig.~\ref{Fig2}(a)]. $\Lambda$ and $\Delta$ can also change due to interactions between hysterons that change the limits of stability $\Gamma^\pm$ and can also cause a bifurcation into a state in which the two minima merge into a single minimum (a supercritical pitchfork bifurcation) \cite{Guckenheimer2013}. This new minimum thus represents an intermediate soft spot configuration that cannot switch [Fig.~\ref{Fig2}(b)]. We will refer to such hysterons as temporarily frozen since their response is purely elastic. 

\begin{figure*}[tb]
\begin{center}
\includegraphics[width=0.90\linewidth]{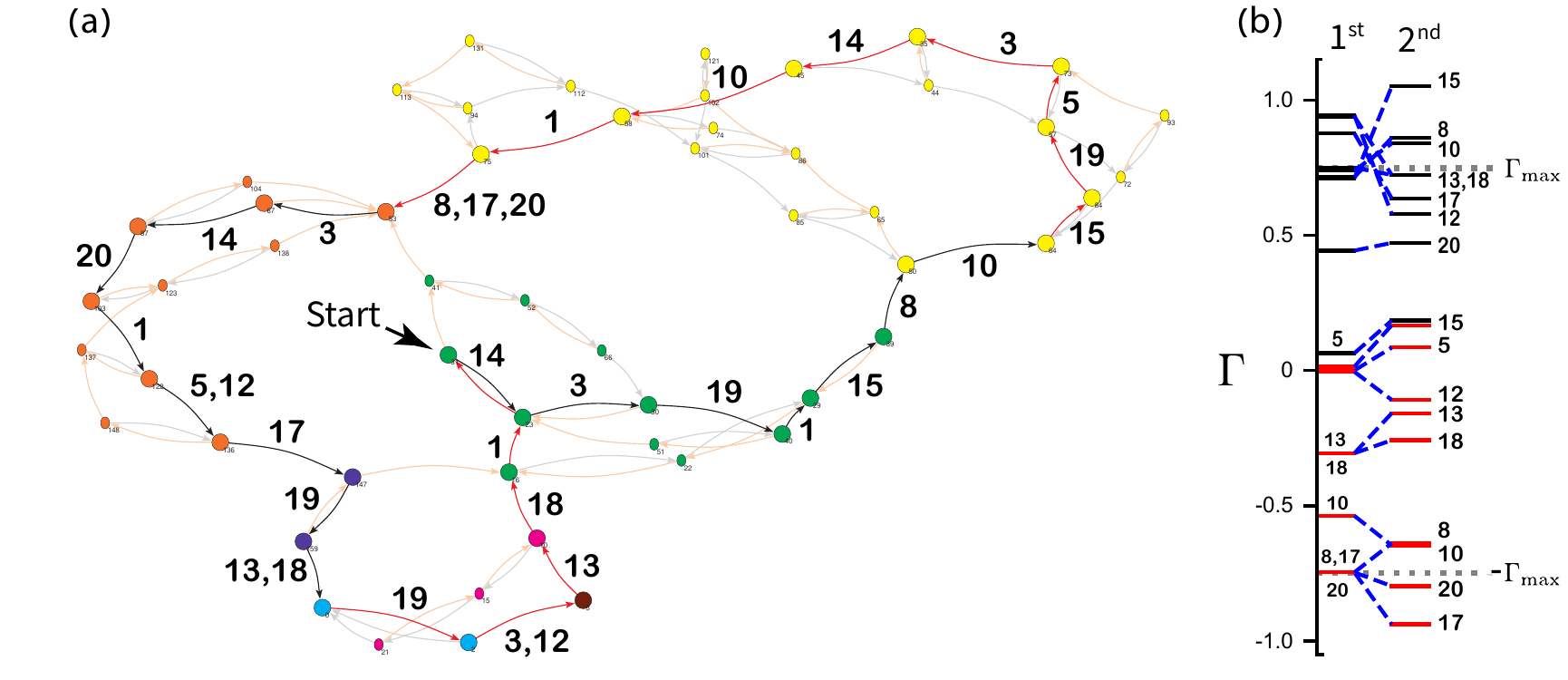}
\caption{(a) An example transition graph of a $T=2$ cycle generated by a system of $N = 20$ interacting hysterons. 
In the cycle shown, 13 soft spots are active, nine of which have a switching pattern of period $2$. An arrow marks the beginning of the first cycle.
(b) The difference in $\Gamma^+_i$ (black) and $\Gamma^-_i$ (red) values of the different soft spots in the two driving cycles. } 
\label{Fig3}
\end{center}
\end{figure*}

\subsection{Adding interactions between hysterons}
To model the conditions in an amorphous solid, we simulate a system of $N$ hysterons, where each hysteron interacts with all of the other hysterons.
As explained above, the interactions between hysterons change the hysteron's switching fields.
As in the amorphous solid, the system is subject to a cyclically varying external field $\Gamma$ with a given maximal strain amplitude $\Gamma_{\max}$.
Each hysteron is assigned a strength coefficient $\epsilon_i$ drawn independently from a uniform distribution ${\cal U}[0.1, 1.0]$, representing a variation in its properties.

In the absence of interactions, the switching fields of each hysteron are given by $\gamma^+_i$ and $\gamma^-_i$, which are drawn independently from the uniform distributions ${\cal U}[0, 1]$ and ${\cal U}[-1, 0]$ respectively. 
We call $\gamma^{\pm}_i$, the \textit{bare} switching strains or strain thresholds. Once interactions are introduced, the thresholds become the \textit{dressed} thresholds  $\Gamma^{\pm}_i$ and are given as
\begin{equation}
	\Gamma^+_{i} = \gamma^+_{i} + \sum_{j\neq i} s_j\epsilon_j\, A^{+}_{i,j},
		\label{eqn:GammaPlusDef}
\end{equation}
and 
\begin{equation}
	\Gamma^-_{i} = \gamma^-_{i} + \sum_{j\neq i} s_j\epsilon_j\, A^{-}_{i,j},
		\label{eqn:GammaMinusDef}
\end{equation}
where $A^{+}$ and $A^{-}$ are the interaction matrices that determine how the upper, respectively lower switching fields of one hysteron change as a result of the state change of other hysterons. 
The matrices $A^{\pm}$ are symmetric and their elements are drawn independently and randomly from a normal distribution with a zero mean ${\cal N}(0, 0.0064)$, which favors the weaker interaction values. In this way, we represent the stronger interaction between the fewer proximate soft spots and weaker interactions with the distant ones.
Note that when a hysteron switches state, the values $\Gamma^\pm_j$ of all the hysterons $j\neq i$ are updated due to the interactions. 

The initial state $s_i=0,1$ of each hysteron is assigned independently and randomly with equal probability.
Choosing the bare $\gamma_i^\pm$ such that $\Gamma^+>0$ and $\Gamma^-<0$ at the start of the simulation run guarantees that the initial configuration is mechanically stable. However, there are no limitations on the values of $\Gamma_i^\pm$ when the system evolves. 

The dynamics is event-driven: in the increasing part of the drive cycle, at each step, the strain $\Gamma$ is incremented to the value of the smallest $\Gamma^+$ from the set of hysterons that are in the state $s_i=0$. When $\Gamma^+ > \Gamma_{\max}$, we set $\Gamma = \Gamma_{\max}$, switch the direction of the strain, and repeat the process where now we switch the strain to the largest $\Gamma^-$ from the set of hysterons with $s_i=1$. We repeat this until $\Gamma^- < -\Gamma_{\max}$ at which point we reverse the drive direction again and so on.

For the iPreisach model, a phenomenon which was not reported before, as much as we are aware,  is that in some cases, after a hysteron changes its state, it can happen that for one of the hysterons $\Gamma^+$ decreases below $\Gamma^-$. We interpret this as the case illustrated in Fig~\ref{Fig2}(b) where the two minima merge. In our algorithm, we assume that this means that the state of this hysteron is frozen and cannot switch until $\Gamma^+$ becomes larger than $\Gamma^-$ again; only then the hysteron is unfrozen and regains its previous state.

To take into account avalanches, after each update of a hysteron $i$, and before proceeding to the next strain threshold, we check whether other hysterons $j\neq i$ become unstable, \textit{i.e.} $\Gamma^+_j < \Gamma$ when $s_j=0$ or $\Gamma^-_j > \Gamma$ when $s_j=1$, in which case we flip these hysterons as well in the same simulation step. In the particle simulations, the dynamics during an avalanche is purely relaxational. In order to be consistent with such dynamics we chose the order of activation during an avalanche such that the hysteron that ``overshot'' $\Gamma$ by the largest extent will be activated first. The order of activation is thus determined according to the \textit{instability} $\Delta\mathcal{E}_j$ of each hysteron, defined as:
\begin{equation}
	\Delta\mathcal{E}_j =
\begin{cases}
	|\Gamma - \Gamma^+_j|,& \text{if } s_j=0 \text{ and } \Gamma^+_j > \Gamma\,, \\
    |\Gamma - \Gamma^-_j|,&  \text{if } s_j=1 \text{ and }\Gamma^-_j < \Gamma\,. \\
\end{cases}
\end{equation}
We then switch the hysteron with the largest $\Delta\mathcal{E}_j$. After every switch we update the values of all $\Gamma^\pm_i$ and repeat the process until all hysterons are stable. 

We next studied the periodic cycles in the iPreisach model. By studying a hysteron model, we can directly follow the changes to the values of $\Gamma^\pm_i$ of the hysterons after each event, which helps us understand the dynamics leading to the strain threshold instability.
To compare the dynamics in the iPreisach model to the particle simulations, we generated a graph using a system of $N=20$ hysterons (comparable to the number of soft spots in Fig~\ref{Fig1}) and found the $T=2$ cycle with the lowest number of active hysterons. The graph of the periodic cycle and the relevant clusters is shown in Fig.~\ref{Fig3} (in this example $\Gamma_{\max} = 0.748$). In Fig.~\ref{Fig3}(a), we see that there are $13$ active hysterons, of which $9$ have period-$2$. Fig.~\ref{Fig3}(b) shows their $\Gamma^+_i$ (black) and $\Gamma^-_i$ (red) values in the first and second driving cycles. We can see that they exhibit the same strain threshold instability that was observed in the $T=2$ cycle that was obtained from particle simulations. 
Hysteron \textit{5} exhibits a different behavior. In the first driving cycle $\Gamma^+_5 < \Gamma^-_5$ and it is therefore temporarily frozen, but $\Gamma^+_5 > \Gamma^-_5$ in the second driving cycle and it can switch states again. To better understand the origin of the threshold oscillations, we first look at a $T=2$ cycle obtained by choosing specific values for $\gamma^\pm$ and the interactions for three hysterons (in~\cite{Keim2021} it was shown that three hysterons form the minimal combination of hysterons that can form a $T=2$ cycle).

\section{Strain threshold in a simple configuration with three hysterons}
\label{section_4}
%
\begin{figure}[tb]
\begin{center}
\includegraphics[width=0.90\linewidth]{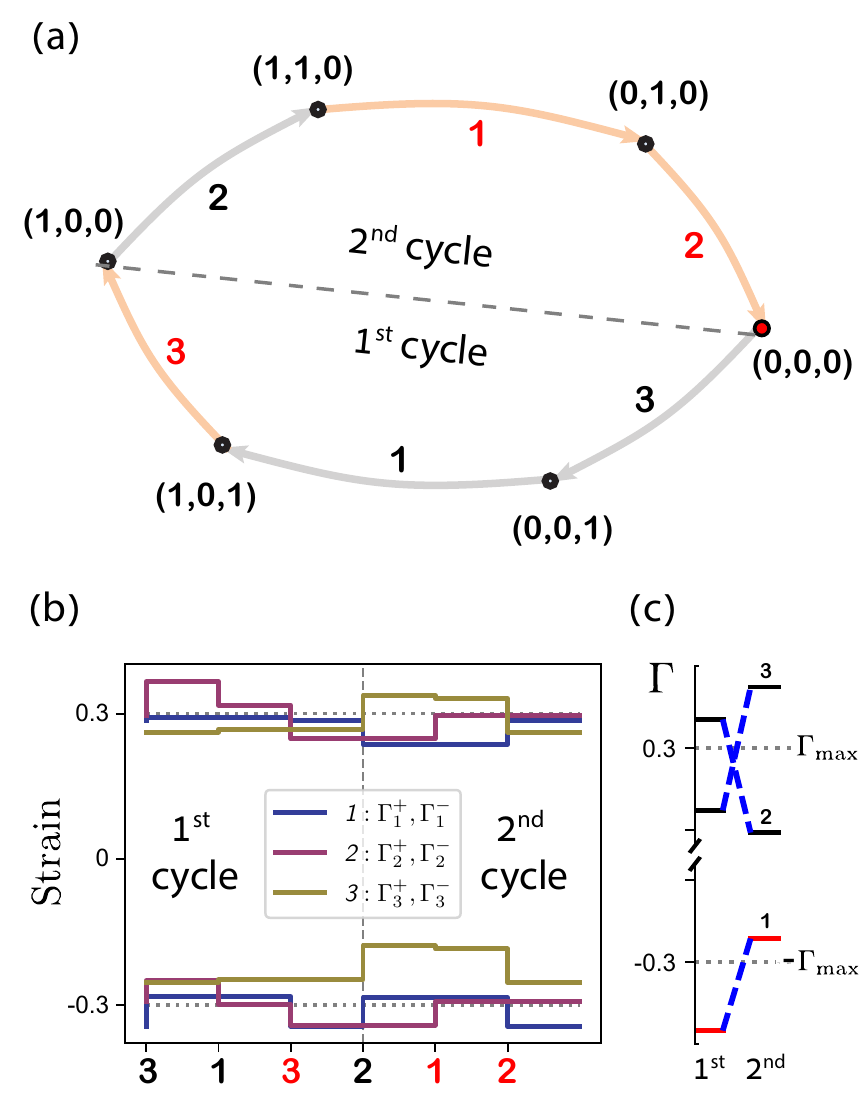}
\caption{ A three-hysteron system whose interactions have been tuned so as to produce a $T=2$ cycle.
(a) Transition graph representation of the $T=2$ cycle. Gray (orange) arrows indicate transitions under increasing (decreasing) strain.
The numbers next to the transitions indicate the hysteron switching its state.
(b) The evolution of the dressed strain thresholds $\Gamma^+_i$ and $\Gamma^-_i$ of hysteron $i = 1, 2$, and $3$ (shown in blue, red, and yellow) and its dependence on the state change of the other hysterons. The labels on the $x$-axis indicated the hysteron changing states from $0 \to 1$ (black label), and $1 \to 0$ (red label).
The horizontal dashed lines indicate the extreme values of the driving,  $\pm\Gamma_{\max}$. (c) A comparison between the values of $\Gamma^+_i$ (black) or $\Gamma^-_i$ (red) of each hysteron in the first and second driving cycle. Horizontal dashed lines indicate  $\pm\Gamma_{\max}$. A hysteron can switch states during a driving cycle, only if $\Gamma^+_i < \Gamma_{\max}$, or $\Gamma^-_i > -\Gamma_{\max}$. Hysteron \textit{3} and \textit{2} can switch only in the first, respectively second driving cycle, resulting in $T =2$ multi-periodic response.
} 
\label{Fig4}
\end{center}
\end{figure}
To understand the origin of threshold oscillations, we study a simple $T=2$ cycle, generated using three interacting hysterons with strain thresholds set to be $\gamma^+ \approx \Gamma_{\max}$ and $\gamma^- \approx -\Gamma_{\max}$ (Fig.~\ref{Fig4}).
The first driving cycle starts from a state $(0,0,0)$. During this cycle hysteron \textit{3} switches two times (once in each segment), hysteron \textit{1} switches once, and hysteron \textit{2} does not switch at all. Therefore, at the end of the first driving cycle the system is in the state $(1,0,0)$. In the second driving cycle, hysteron \textit{2} switches twice, hysteron \textit{1} switches once, and hysteron \textit{3} does not switch at all, which brings the system back to the $(0,0,0)$ state. When looking at the dressed strain thresholds $\Gamma^\pm_i$ of the three hysterons, we can see that all three are period-$2$ hysterons that exhibit the kind of threshold oscillations that were also observed in particle simulations. We also see that each hysteron switches in different segments of the cycle, and therefore, enabling the threshold oscillations of the other hysterons. This means that the interactions cause a collective periodic behavior of out-of-phase threshold oscillations.

More specifically, During the first driving cycle $\Gamma^+_1$ and $\Gamma^+_3$ are smaller than the maximal applied strain $\Gamma_{\max}$ which causes them to switch, while $\Gamma^+_2$ is larger than $\Gamma_{\max}$ and for that reason it does not switch during the first cycle. 
In this example it is clear what is causing the oscillations: once $s_3=1$, $\Gamma^-_1$ changes from its bare value $\gamma^-_1$ to its dressed value:
\begin{equation}
\Gamma^-_1 = \gamma^-_1 + \epsilon_3\, A^{(-)}_{1,3} < -\Gamma_{\max}\,
\end{equation}
and for that reason \textit{1} remains in $s_1=1$ until the second driving cycle. 
When \textit{3} switch back to $s_3=0$, the \textit{1} negative threshold goes back to its bare value $\gamma^-_1$ which is larger than 
$-\Gamma_{\max}$ and thus can switch back to $s_1=0$ during the second driving cycle.
Similarly, when $s_3=1$, the strain threshold of \textit{2}, is pushed to a value that is larger than $\Gamma_{\max}$ and thus $s_2=0$ during the first driving cycle. But when $s_1=1$ and $s_3=0$, $\Gamma^+_2$ is pushed to a value that is smaller than $\Gamma_{\max}$ and can therefore switch in the second driving cycle. A detailed set of conditions for this type of behavior is given in Appendix~D.

The manner in which interactions cause strain threshold oscillations in this simple realization of the model is easy to understand. In a more realistic realization of the model, such as the one shown in Fig~\ref{Fig3}, the mechanism is more complex. As we will see in the next section, in this case, multi-periodic dynamics of specific hysterons is some times a result of cooperative dynamics involving soft spots that are not multi-periodic.
Note also that in order to have $T=2$ cycles, some interactions must be stabilizing, since a system with purely destabilizing interactions can not generate multi-periodic cycles (see Appendix~\ref{AppE}).
\begin{table*}[tb]
\begin{tabular}{ccccccccccccccccccccccccc}
\hline
\multicolumn{1}{c}{\multirow{2}{*}{\textbf{Period}}} & \multicolumn{1}{c}{\multirow{2}{*}{\begin{tabular}[c]{@{}c@{}}\textbf{Strain} \\ \textbf{direction}\end{tabular}}} & \multicolumn{23}{c}{\textbf{Soft spot index}}                                                                           \\
\multicolumn{1}{c}{}                        & \multicolumn{1}{c}{}                                                                             & \cellcolor{yellow!50}1 & \cellcolor{yellow!50}9 & 11 & 14 & \cellcolor{yellow!50}16 & 21 & \cellcolor{green!25}23 & 27 & 33 & 41 & 43 & 44 & 50 & 55 & 59 & 69 & 73 & 79 & 85 & 87 & 89 & 91 & \cellcolor{yellow!50}93 \\ \hline
\multirow{2}{*}{$T=1$}                         & $\nearrow$                                                                                               & \cellcolor{yellow!50}1 & \cellcolor{yellow!50}1 & 1  & 1  & \cellcolor{yellow!50}1  & 1  & \cellcolor{green!25}1  & 1  & 1  & 1  & 1  & 1  & 1  & 1  & 1  & 1  & 1* & 1  & 1* & 1  & 1  & 1  & \cellcolor{yellow!50}1  \\
                                            & $\searrow$                                                                                             & \cellcolor{yellow!50}1 & \cellcolor{yellow!50}0 & 0  & 0  & \cellcolor{yellow!50}0  & 0  & \cellcolor{green!25}1  & 0  & 0  & 0  & 0  & 0  & 0  & 0  & 0  & 0  & 0  & 0  & 0  & 0  & 0  & 0  & \cellcolor{yellow!50}0* \\ \hline
\multirow{4}{*}{$T=2$}                         & $\nearrow$                                                                                               & \cellcolor{yellow!50}1 & \cellcolor{yellow!50}1 & 1  & 1  & \cellcolor{yellow!50}0  & 1  & \cellcolor{green!25}0  & 1  & 1  & 1  & 1  & 1  & 1  & 1  & 1  & 1  & 1  & 1  & 1  & 1  & 1  & 1  & \cellcolor{yellow!50}1  \\
                                            & $\searrow$                                                                                             & \cellcolor{yellow!50}0 & \cellcolor{yellow!50}0 & 0  & 0  & \cellcolor{yellow!50}0  & 0  & \cellcolor{green!25}0  & 0  & 0  & 0  & 0  & 0  & 0  & 0  & 0  & 0  & 0  & 0  & 0  & 0  & 0  & 0  & \cellcolor{yellow!50}1* \\
                                            & $\nearrow$                                                                                               & \cellcolor{yellow!50}1 & \cellcolor{yellow!50}0 & 1  & 1  & \cellcolor{yellow!50}1  & 1  & \cellcolor{green!25}0  & 1  & 1  & 1  & 1  & 1  & 1  & 1  & 1  & 1  & 1  & 1  & 1* & 1  & 1  & 1  & \cellcolor{yellow!50}1  \\
                                            & $\searrow$                                                                                             & \cellcolor{yellow!50}1 & \cellcolor{yellow!50}0 & 0  & 0  & \cellcolor{yellow!50}0  & 0  & \cellcolor{green!25}0  & 0  & 0  & 0  & 0  & 0  & 0  & 0  & 0  & 0  & 0  & 0  & 0  & 0  & 0  & 0  & \cellcolor{yellow!50}0* \\ \hline
\end{tabular}
\caption{\label{tab:state_table} The system's state ${\bf S}$ after each segment of the driving cycle ($\nearrow$ and $\searrow$ for increasing and decreasing strain respectively). The $T=2$ cycle begins after hysteron \textit{23} is activated. In yellow are the period-$2$ hysterons, green marks hysteron \textit{23}, which is fixed at $s_{23}=1$ in the $T=1$ cycle. The star superscripts mark the frozen hysterons at the time of the measurement due to $\Gamma^+_i$ being smaller than $\Gamma^-_i$.}
\end{table*}
\begin{table*}[tb]
\begin{tabular}{ccccccccccccccccccccccc}
\hline
\multicolumn{1}{c}{\textbf{Period}} & \multicolumn{1}{c}{\begin{tabular}[c]{@{}c@{}}\textbf{Strain}  \textbf{direction}\end{tabular}} & \multicolumn{21}{c}{\textbf{Soft spot switching order}}                                                        \\ \hline
\multirow{2}{*}{$T=1$}        & $\nearrow$                                                                              & 85 & 59 & 44 & 73 & 79 & 33 & 50 & 89 & 91 & 11 & 14 & 43 & 55 & 87 & 9  & 41 & 93 & 69 & 21 & 27 & 16 \\
                           & $\searrow$                                                                            & 9  & 41 & 73 & 55 & 43 & 87 & 93 & 89 & 11 & 79 & 50 & 14 & 44 & 91 & 59 & 85 & 27 & 69 & 33 & 27 & 16 \\ \hline
\multirow{4}{*}{$T=2$}        & $\nearrow$                                                                              & 85 & 59 & 44 & 79 & 73 & 33 & 50 & 89 & 91 & 11 & 14 & 43 & 55 & 87 & \cellcolor{yellow!50}9  & \cellcolor{red!20}41 & \cellcolor{orange!25}93 & \cellcolor{red!20}69 & \cellcolor{red!20}21 & 27 &    \\
                           & $\searrow$                                                                              & 41 & \cellcolor{yellow!50}9  & 55 & 73 & 43 & 87 & 11 & 85 & 14 & 50 & 89 & 79 & 44 & 27 & 59 & 91 & 69 & 33 & 21 & 1  &    \\
                           & $\nearrow$                                                                              & 1  & 85 & 59 & 44 & 73 & 79 & 33 & 50 & 89 & 11 & 91 & 43 & 14 & 55 & 87 & \cellcolor{red!20}69 & \cellcolor{red!20}21 & \cellcolor{red!20}41 & 16 & 27 &    \\
                           & $\searrow$                                                                              & 41 & 73 & 55 & 43 & 87 & \cellcolor{orange!25}93 & 89 & 11 & 79 & 50 & 14 & 44 & 91 & 59 & 85 & 27 & 69 & 33 & 21 & 16 &    \\ \hline
\end{tabular}
\caption{\label{tab:activation_order} The switching order of the hysterons in the $T=1$ and $T=2$ cycles (left top right), corresponding to Table~\ref{tab:state_table}. The colored cells mark how hysteron \textit{93} (orange) causes threshold oscillations in hysteron \textit{9} (yellow) via the cascade sequence of hysterons \textit{41},  followed by \textit{69} and \textit{21} (red) which in the second driving cycles is scrambled, \textit{69}, followed by \textit{21} and \textit{41}. This cascade-scrambling and the resulting threshold oscillation has been visually highlighted in Fig.~\ref{Fig5}(c).
}
\end{table*}
%
\section{$T>1$ in a multi-hysteron system}
\label{section_5}
%
\begin{figure*}[tb]
\begin{center}
\includegraphics[width=0.90\linewidth]{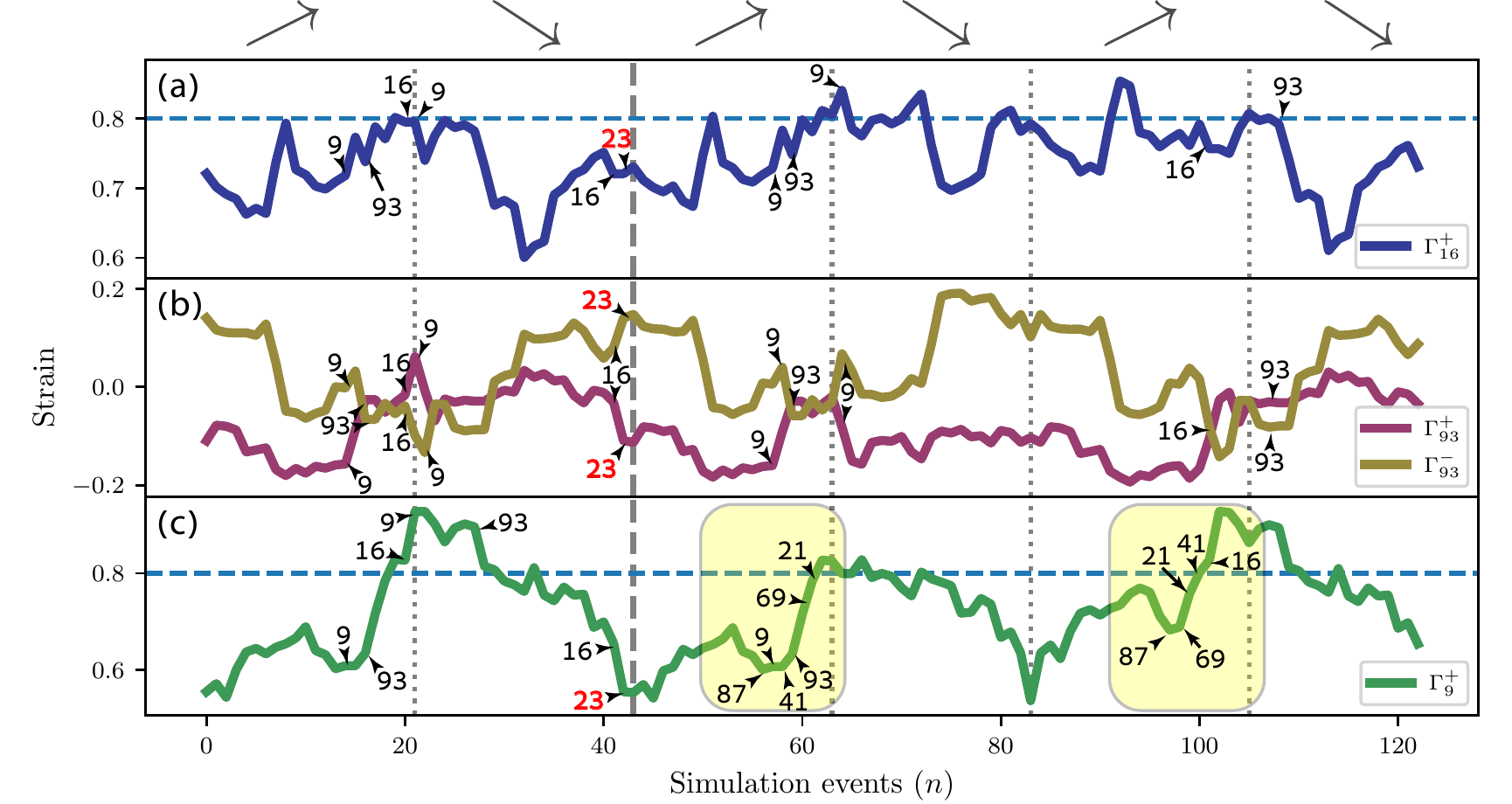}
\caption{Event graph showing the evolution of  $\Gamma^+_i$ and $\Gamma^-_i$ of hysterons $i =$ \textit{16} (a), \textit{93} (b), and  \textit{9} (c) from one switching event to the next. Here $\Gamma_{\max} = 0.8$ (dashed horizontal lines), the switching of hysteron \textit{23} is marked by a dashed vertical line, and the dotted vertical lines show the different segments of the driving cycle, alternating from $\nearrow$ in the left to $\searrow$ in the right. The numbers of selected hysterons are above or below the line. The complete switching order is given in Table~\ref{tab:activation_order}. The yellow regions in (c) highlight the cascade effect pushing $\Gamma^+_{9}$ beyond $\Gamma_{\max}$.
} 
\label{Fig5}
\end{center}
\end{figure*}
As we saw in the previous section, the mechanism leading to threshold oscillations and $T=2$ in a 3-hysteron system, is relatively simple.
However, in a many-hysteron system, which is typically the case in an amorphous solid subject to oscillatory shear, the mechanism can be more complex and can  involve the cooperative, time-ordered dynamics of several hysterons. 
We therefore consider a system with $N=100$ hysterons that is driven by cyclic shear at some strain amplitude $\Gamma_{\max}$. In particular, we focus on a realization of this multi-hysteron system that gives rise to multi-periodic response, \textit{i.e.} $T > 1$. 
We are interested in characterizing how a transition from periodic response with $T = 1$ to multi-periodic response sets in. 
In order to do so, we found a $T=3$ cycle, and starting from its zero-strain configuration, we \textit{knock-out} all $N$ hysterons by manually freezing them, \textit{i.e.} by fixing their states $s_i$ to the state at the beginning of the cycle.
We then choose a hysteron $i$ at random and activate it, by which we mean that we are allowing it to switch freely between states. 
We continue activating the hysterons one by one and at random.
After each activation, we let the system evolve to attain cyclic response. The cycles found after the activation of the first two hysterons must necessarily have $T=1$, as multi-periodic response requires at least three hysterons \cite{Keim2021, Lindeman2021}.
Since the full $N$-hysteron system exhibits multi-periodic response, at some point the activation of another hysteron must lead to a response with $T > 1$. 
In the example considered here, multi-periodic response sets in for the first time after $29$ hysterons have been activated: with the activation of 
hysteron \textit{23} a multi-periodic cycle with $T=2$ is attained.

Adding \textit{23} is therefore the \textit{tipping point}, allowing the system of hysterons to switch from $T=1$ to $T=2$. We next compare this $T=2$ cycle to the $T=1$ cycle reached just before hysteron \textit{23} was activated. Table~\ref{tab:state_table} compares the switching-patterns of the hysterons in the $T=1$ and $T=2$ cycles.  We see that all of the hysterons that were active in the $T=1$ cycle are also active in the new $T=2$ cycle. Note that although hysteron \textit{23} triggered the transition to the $T = 2$ cycle, once it has been activated it switches from $1$ to $0$, and then remains frozen in state $0$ and hence is not active in the $T=2$ cycle anymore (note that some active hysterons, {\it i.e.} hysterons that are allowed to switch, do not actually switch during the cycle). 
In fact, the sets of hysterons switching in the $T=1$ and $T=2$ cycles are almost identical and differ only by the participation of hysteron \textit{1}, the only active hysteron not switching in the $T=1$ cycle but switching in the $T=2$ cycle. However, its switching pattern is rather peculiar. To see this, we depict in Table~\ref{tab:activation_order} the \textit{switching-order} of the hysterons active in
the $T = 1$ and $T=2$ cycles, \textit{i.e.} the order in which the individual hysterons change state during the different segments of these cycles.   
Hysteron \textit{1} is the last hysteron to switch its state from $0$ to $1$ during the $\nearrow$ segment of the first driving cycle, but also the first one to switch it back to $0$ when the strain starts to decrease in the second driving cycle. Thus its switching pattern 
does not affect the dynamics of the other hysterons. 
We will therefore ignore hysteron \textit{1} in the following. 
With hysteron \textit{1} excluded, the sets of hysterons active in the $T=1$ and $T=2$ cycles are identical.
\begin{figure*}[tb]
\begin{center}
\includegraphics[width=0.90\linewidth]{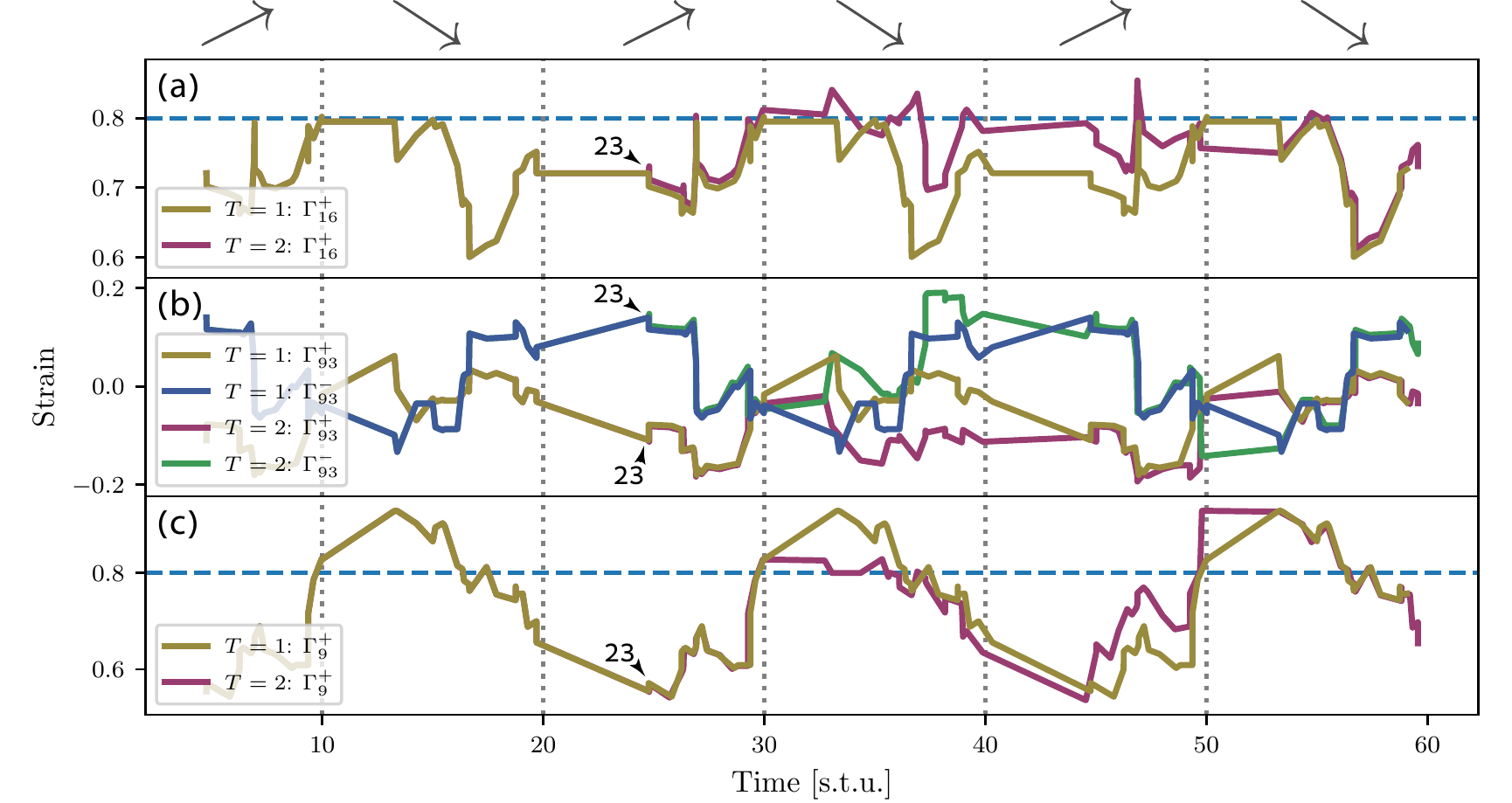}
\caption{Time evolution of the strain thresholds $\Gamma^+_i$ and $\Gamma^-_i$ for hysterons $i = $ \textit{16} (a), \textit{93} (b), and \textit{9} (c). The period of the driving is $20$ in simulation time units (s.t.u.) and the dashed vertical lines mark the $\nearrow$- and $\searrow$-segments  (in this order from left to right) of each driving cycle. The driving amplitude $\Gamma_{\max} = 0.8$ has been indicated by the dashed horizontal lines. 
In each panel we compare the evolution of the strain thresholds when \textit{23} is manually frozen, and the cyclic response has period $T=1$ (yellow and blue curves), and when hysteron \textit{23} is activated in the second driving cycle $20 \le t \le 40$  (red and green curves). Following its activation, \textit{23} switches states near $t = 25$, and the system settles into a $T = 2$ multi-periodic response at $t \ge 40$. The curves show a temporary divergence of the switching fields for the $T = 1$ and $T = 2$ response after  the state change of \textit{23}, before they tend to move together, albeit with an offset of strain values. 
}
\label{Fig6}
\end{center}
\end{figure*}

We can think of the triggering effect of hysteron \textit{23}  as a perturbation to the strain thresholds $\Gamma^{\pm, 1T}_i$ of the hysterons active in the $T=1$ cycle. From Eq.~(\ref{eqn:GammaPlusDef}) and Eq.~(\ref{eqn:GammaMinusDef}) we find that the strain thresholds $\Gamma^{\pm, 2T}_{i}$ of the $i$th hysteron in the $T=2$ cycle are given in terms of $\Gamma^{\pm, 1T}_i$ as
\begin{equation}
\Gamma^{\pm, 2T}_{i} = \Gamma^{\pm, 1T}_{i} - \epsilon_{23}A_{i,23}^{\pm}\,.
\end{equation}
This shift in the strain thresholds causes three hysterons, \textit{16}, \textit{93}, and \textit{9}, to have a switching pattern with period $T=2$. In Fig.~\ref{Fig5}(a), we can see the emerging switching pattern: \textit{23} causes the activation threshold of \textit{16}
to shift so that it does not switch during the first forcing cycle. This causes \textit{93} to stay \textit{frozen} and not switch during an entire cycle, which then causes the strain threshold of \textit{9} to shift such that it also does not switch. 
This situation is similar to the three-hysteron system where one hysteron, switching or not switching, affects the switching pattern of another hysteron.
However, the mechanisms that cause a hysteron not to switch are not always simple threshold oscillations. This can be seen in the dynamics of \textit{93} and \textit{9}. Hysteron \textit{93} does not switch in the second forcing cycle because $\Gamma^+_{93}$ and $\Gamma^-_{93}$ are shifted such that $\Gamma^+_{93} < \Gamma^-_{93}$ during the following forcing cycle [Fig~\ref{Fig5}(b)], and \textit{93} is thus frozen in the sense discussed in Fig \ref{Fig2}(b) and the introduction.

In the dynamics of \textit{9}, shown in Fig~\ref{Fig5}(c) and Table~\ref{tab:activation_order}, we see yet another mechanism causing a hysteron to have $T=2$. 
In the first forcing cycle, hysteron \textit{9} switches directly after \textit{87} and before \textit{93}. However, because \textit{93} did not switch in the $\searrow$ segment, in the second forcing cycle, \textit{69} and \textit{21} switches before \textit{41}, which then causes $\Gamma^+_9$ to become larger than $\Gamma_{\max}$, and hysteron \textit{9} does not switch at all during the second forcing cycle. Therefore, the fact that \textit{93} stayed frozen caused a cascade of events that indirectly causes $\gamma^+_{9}$ to become temporarily larger than $\Gamma_{\max}$. 
We observe that the switching order of \textit{69}, \textit{41} and \textit{21} changes periodically between the first and second $\nearrow$ segments. The different state of hysteron \textit{93} therefore {\it scrambles} the switching order of these three hysterons, where in the first $\nearrow$ segment these hysterons switch in the order \textit{41, 69, 21}, while in the second, the switching order is \textit{69, 21, 41}. The combined effect of the cascade-scrambling thus causes the threshold of hysteron \textit{9} to oscillate around $\Gamma_{\max}$ with period $T=2$.

To get a better understanding of the difference between the $T=1$ and $T=2$ cyclic response, we plotted in Fig~\ref{Fig6} the same strain thresholds $\Gamma^\pm_i$ that were shown in Fig~\ref{Fig5}, but this time against simulation time, so that events in the two cycles that occur at the same time, necessarily must also occur at the same driving strain $\Gamma$. This allows us to compare the differences in $\Gamma^\pm_i$ of the hysterons whose switching pattern changes to period-$2$.  
Here we can clearly see how the strain thresholds in the $T=2$ cycle move away from their values in the $T=1$ cycle, due to the instability discussed above, but then, move back again towards the $T=1$ cycle values after about one driving cycle. This also demonstrates how the period-2 hysterons grow the relatively small perturbations to $\Gamma^\pm_i$ into a significant shift, giving rise to the threshold oscillations. It is apparent that because the perturbation is relatively weak, the strain thresholds return to their values in the $T=1$ cycle when the states of the period-2 hysterons return to their states at the time of the perturbation.

\section{Evidence for temporary freezing and the cascade-scrambling effect in the particle simulations}
\label{sec:comparePreisachParticle}
The cascade-scrambling effect as a form of indirect interaction mechanism for multi-periodic response is more complex than the simple sequential threshold oscillations observed in the three-hysteron system, as it involves the alteration of the switching order of several hysterons. In the simulations of particles systems and elastoplastic models, it is well-known that with increasing driving amplitude, the resulting periodic response involves an increasingly larger number of active soft spots. Thus a priori, a cascade-scrambling effect will be feasible only at sufficiently large driving amplitudes. At the same time, the simulations show that with increasing driving amplitude, a multi-periodic response becomes increasingly more likely. On the other hand, 
estimates for the likelihood of multi-periodic $T=2$ response in systems with three interacting hysterons and randomly chosen interactions, as carried out in \cite{Keim2021}, reveal that such a response is extremely unlikely. Thus the effect of cascade-scrambling may explain the observed abundance of multi-periodic response in numerical simulations at large driving amplitudes.   

We therefore revisit the $T =2$ cycle observed in our particle simulation to see whether it contains soft spots whose switching pattern resembles those of temporarily frozen ones, and to seek evidence of the presence of a cascade-scrambling effect. In the particle simulations, even though we are unable to follow the values of $\Gamma^{\pm}_i$ for each of the soft spots at all times, we can indeed infer switching patterns and changes in $\Gamma^{\pm}_i$ that indicate the existence of these two mechanisms,  i.e. temporary freezing of a hysteron and cascade-scrambling. 
In particular, from Fig.~\ref{Fig1} we see that the first time that  \textit{10} switches, $\Gamma^+_{10}$ is negative and lower than $\Gamma^-_{10}$. We also find that $|\Gamma^\pm_{10}|$ is much smaller than $\Gamma_{\max}$, which means that it is not likely to oscillate around $\Gamma_{\max}$. This behavior resembles the dynamics of the strain thresholds of a temporarily frozen hysteron.
In fact, in the interacting hysteron system, if we had observed the strain thresholds of the temporarily frozen hysteron only when it is active and changing states, the evolution of these thresholds would be similar to that of the strain thresholds of soft spot \textit{10} as well as \textit{2}. Appendix \ref{AppC} contains several more examples of hysterons temporarily frozen during a $T=3$ cycle obtained from particle simulations. Moreover, in Fig.~\ref{Fig1} we also observe changes in the switching order of soft spots that may thus be indicative of cascade-scrambling: while \textit{3} switches before \textit{4} in the $\searrow$ segment of the first cycle, \textit{4} and \textit{5} switch before \textit{3} in the $\searrow$ segment of the second cycle. This suggests that soft spot \textit{5} may exhibit threshold oscillations due to indirect interactions with another period-$2$ soft spot via the alteration of the switching order of soft spots \textit{3} and \textit{4}.

Another mechanism observed in particle simulations is overlapping soft spots. The phenomenon appears in Fig.~\ref{Fig1}(a) as overlapping ellipses, which means that two (or more) soft spots share particles. One implication is the existence of conditional switching patterns, meaning that one soft spot cannot switch before an overlapping one does. For example, we see that soft spots \textit{7} and \textit{9} overlap and indeed \textit{7} switches before \textit{9} in the $\searrow$ segment of the first cycle and then switches after 9 in the $\nearrow$ segment of the second cycle. This behavior was also found in the $T=3$ cycle discussed in Appendix~\ref{AppC}. In the current simulation results we did not find any indication that overlap affects multi-periodicity. However we suspect that overlaps could play a role in the dynamics of $T>3$ cycles, which may require modification of the iPreisach model.

\section{Discussion}
\label{sec:discussion}
The appearance of periodic and multi-periodic response and reversible plasticity in sheared amorphous solids has attracted significant interest. 
Here we focused specifically on multi-periodic response since while it was shown that simple models of interacting hysterons reproduce multi-periodic responses, the exact manner in which interactions lead to such response is still unclear. Furthermore, a direct connection between the multi-periodic response in these systems and in amorphous solids was not established.

We developed an algorithm based on a graph representation of the configuration space to identify all the soft spots that participate in a $T=2$ cycle extracted from particle simulations. We found that some soft spots return to their initial states after one driving cycle, while others repeat after two driving cycles and thus generate  multi-periodicity. We also found that the multi-periodicity of some of these soft spots stems from oscillations of their switching strain values around the strain amplitude, which causes them to switch only during alternating forcing cycles. 
Finally, we identified other soft spots exhibiting period-2 dynamics, which did not seem  to stem from such oscillations. 

To clarify how soft spot interactions lead to multi-periodic behavior, we modeled the system as a set of interacting hysterons that switch between two states in a hysteretic manner and represent individual soft spots. As was recently demonstrated \cite{Keim2021, Lindeman2021}, such a system also exhibits multi-periodic behavior when subject to oscillatory driving. To understand the origin of multi-periodic dynamics in this model, we compared the dynamics of a $T=1$ cycle to a $T=2$ cycle obtained from the former by adding one hysteron. We showed that the same hysterons were switching in both cycles, but in the $T=2$ cycle, some hysterons repeated after one driving cycle, while others repeated after two cycles. Unlike the direct interaction mechanism that leads to sequential oscillations of the strain thresholds in a simple three-hysteron system, in a system with a large number of hysterons, we have shown that some of the period-$2$ hysterons exhibit threshold oscillations as a result of cascade-scrambling, \textit{i.e.} changes in the switching order of other hysterons. Furthermore, we suggested another dynamic threshold oscillations mechanism, the freezing mechanism, that causes individual hysterons to develop period-2 due to a temporary loss of their ability to switch.

The nature of the mechanisms that we uncovered has significant implications for the understanding, the modeling, and the uses of plasticity in amorphous solids. It has long been recognized that interactions between soft spots play an important role in the plasticity of amorphous solids. Up until now, the focus has been on the role that interactions play in the size and duration of avalanches, \textit{i.e.} plastic events in which the switching of one soft spot induces a displacement field that then causes other soft spots to switch. However, the order in which these soft spots switch was not thought to play an important role in the dynamics. Here we have shown that changes in the switching order of soft spots is also significant for the dynamics, at least under periodic forcing. Therefore, a theoretical or numerical model of the dynamics of amorphous solids under oscillatory shear should take into account switching order effects. Besides the effects of soft spot interactions on the order of switching, we have also found evidence suggesting that interactions may cause a soft spot to temporarily not switch, which may also be important for understanding plasticity in amorphous solids.

Both effects may be useful for manipulating memory encoding and readout in amorphous solids.
Since the interacting hysteron model is rather generic, our results may also be relevant to other mechanical systems. For example, scrambling effects have been observed in corrugated elastic sheets \cite{bense2021} and therefore, it is possible that manipulating such effects can allow the creation and manipulation of cycles with periods larger than one also in these or related systems. 
\bigbreak
{\bf Acknowledgments} I.R. and A.S. were supported by the Israel Science Foundation through Grant No.~1301/17.
MM was supported by the Deutsche Forschungsgemeinschaft (DFG, German Research Foundation) under Projektnummer 398962893, the Deutsche Forschungsgemeinschaft (DFG, German Research Foundation) - Projektnummer 211504053 - SFB 1060, and by the Deutsche Forschungsgemeinschaft (DFG, German Research Foundation) under Germany’s Excellence Strategy - GZ 2047/1, Projekt-ID 390685813.

\normalem

\appendix

\section{Particle simulations}
\label{App0}
We simulated $1024$ particles in two-dimensions interacting with a radially-symmetric potential described in \cite{Lerner2009, Regev2013} and integrated with a leap-frog solver \cite{Allen2017}. $50$\% of the particles were $1.4$ times larger than the other $50$\% to prevent crystallization. To prepare the initial configurations used to construct the $t$-graphs, we performed particle simulations with an Andersen thermostat that kept the system at a high temperature $T=1$ (in Lennard-Jones energy units) in a liquid state. The simulation ran for $20$ simulation time units, after which the temperature was reduced to $T=0.1$ and was run for another $50$ simulation time units. The final configuration of this run was then quenched to zero temperature using the FIRE minimization algorithm \cite{Bitzek2006}. 
The resulting configuration was then used as the initial configuration to construct the $t$-graph. Deformations were applied quasistatically, such that at each simulation step, the sample was sheared by $10^{-4}$ using the periodic Lees-Edwards boundary conditions \cite{Lees1972} that was followed by minimizing the energy using the FIRE algorithm. Starting from any configuration, the strain was increased until a plastic event occurred. This protocol was used to construct the $t$-graph as described below.

\section{Constructing a $t$-graph}
\label{AppA}
To construct the graph, we start with a configuration that is stable at some strain $\Gamma$ (in this case, a state that is part of the periodic cycle of interest), and identify its strain thresholds: the $\Gamma^+$ and $\Gamma^-$ at which a plastic event which includes one or more soft spots switching states. 
All the configurations in the range $\Gamma^- < \Gamma < \Gamma^+$ are represented by a single mesostate $O$, represented as a vertex on the graph. The plastic events at $\Gamma^+[O]$ and $\Gamma^-[O]$ lead to new configurations that are part of mesostates $A$ (that is reached after $\Gamma^+[O]$) and $B$ (that is reached after $\Gamma^-[O]$) that have different strain thresholds $\Gamma^+[A]$, $\Gamma^-[A]$ and $\Gamma^+[B]$, $\Gamma^-[B]$ respectively. The mesostates $A$ and $B$ are represented as two separate vertices. 
The plastic transition leading from $O$ to $A$ is thought of as an operator $\Up$ (for an ``up'' transition), that represents a transition due to an increase in the strain:
\begin{equation}
A = \Up O\,,
\end{equation}
and is represented in the $t$-graph as a black (or gray) arrow. The plastic transition from $O$ to $B$ is represented by the operator $\Dn$ (for a ``down'' transition), which represents a transition due to a decrease in the strain:
\begin{equation}
B = \Dn O\,,
\end{equation}
and is shown as a red (or orange) arrow in the $t$-graph.
This process is repeated iteratively for each new mesostate $M$, where plastic transitions at $\Gamma^+[M]$ and $\Gamma^-[M]$ can lead to new mesostates or back to mesostates that were encountered before. When this happens, a graph loop is closed.
Detailed explanations and examples are given in~\cite{Mungan2019} and its supplementary material; 
here we have followed the notation used in related studies~\cite{Mungan2019, Mungan2019a, Mungan2019b, Regev2021, Hecke2021, bense2021, Keim2021, Lindeman2021}.

\section{Soft spot identification methodology}
\label{AppB}
The basic cycles that we investigate are cycles with a perfect or near-perfect $\ell$RPM as defined in the text. This guarantees 
that any soft spot that is switched on $s_i = 0\rightarrow 1$ is switched off  $s_i = 1\rightarrow 0$ in another plastic event that is part of the same cycle or one of its sub-cycles. 
By using this fact and appropriate book-keeping, together with a method to compare the displacement fields ${\bf v}$ of different plastic events, allows us to systematically identify the different soft spots.
To identify whether two different displacement fields represent the same plastic event, we compare ${\bf v}$, the $2N$ vector of displacements during one plastic event ($N$ is the number of particles):
\begin{equation}
{\bf v} = \{v_x^1,v_y^1,v_x^2,v_y^2,....,v_x^N,v_y^N\}
\end{equation}
to the vector ${\bf u}$ of the other plastic event. Here $(v_x^i,v_y^i)$ is the displacement of the $i$th particle. To compare the fields ${\bf v}$ and ${\bf u}$, we calculate the scalar product of the normalized displacement vectors $\hat{{\bf v}}$ and $\hat{{\bf u}}$:
\begin{equation}
\hat{{\bf v}}\cdot\hat{{\bf u}} \equiv \frac{1}{N}\sum_{i=1}^N (\hat{v}_x^i\hat{u}_x^i + \hat{v}_y^i\hat{u}_y^i)\,.
\end{equation}
If the plastic events involve the same set of soft spots, the result will be $\sim 1$ or $\sim -1$ (we used a threshold of $0.98$), where the latter corresponds to cases when the compared fields are of the same soft spots in transitions of increasing and decreasing strain (in the case of a single soft spot this is a case where we are comparing the displacement fields generated by the same soft spot $i$ in a transition $s_i=0 \to 1$ to the displacement field generated by the same soft spot in the transition $s_i=1 \to 0$). In plastic events where more than one soft spot are switched (an avalanche) we have to use special care. For example, say that we have situation where a transition that results from an increase in the strain, causes soft spots \textit{1} and \textit{2} to switch, and in a transition that results when the strain is decreased, soft spot \textit{1} appears together with \textit{3}. 
The displacement field associated with the transition of soft spot \textit{1} can then be found by subtracting the displacement field of soft spot \textit{2} or soft spot \textit{3} from the displacement field of the respective plastic events, assuming that the fields can be linearly superimposed (more complex combinations of soft-spots can also be decomposed this way). This assumption can break down when the soft spots are too close to each other, but still worked in most of the cases that we have studied. When this assumption is not satisfied we have to intervene manually.

\section{particle simulation graph of a $T=3$ cycle}
\label{AppC}
\begin{figure*}[tb]
\begin{center}
\includegraphics[width=0.90\linewidth]{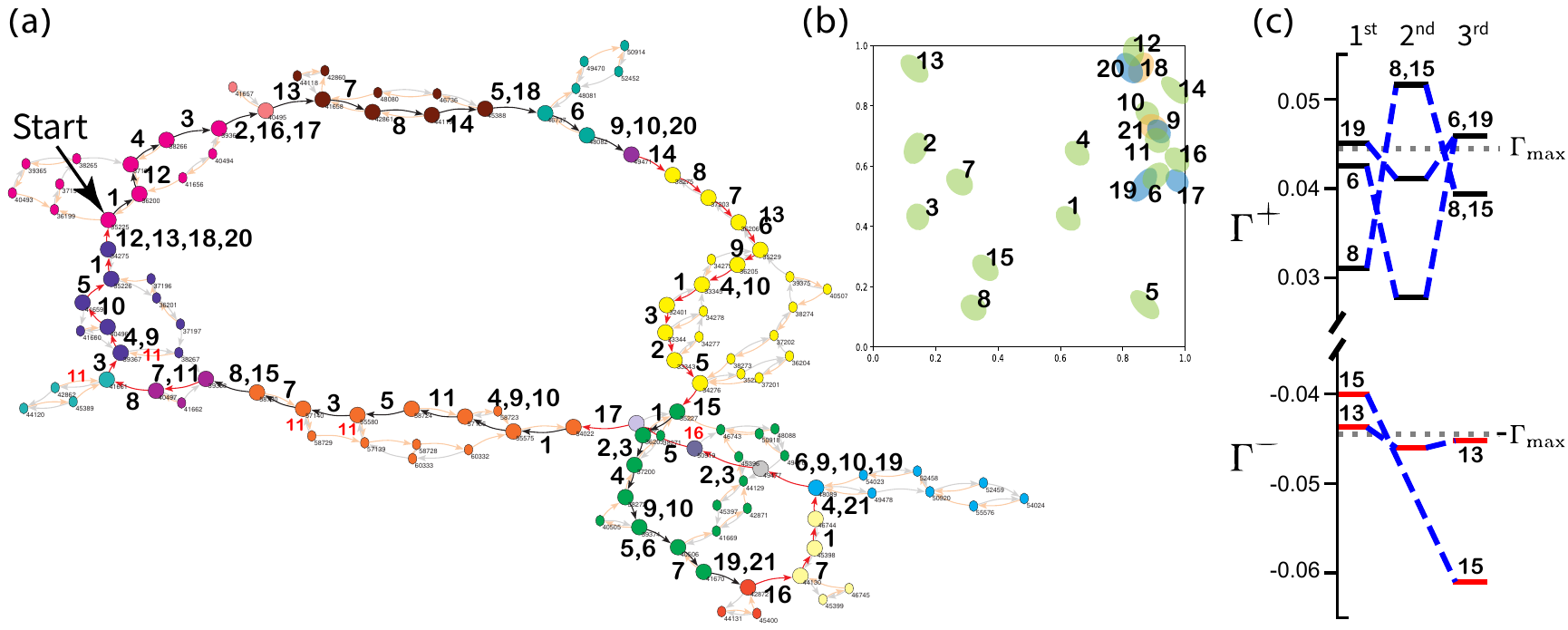}
\caption{(a) A graph representation of a $T=3$ cycle generated by a particle simulation. Black and gray arrows mark the transitions under increasing strain, while transitions as a result of decreasing strain are shown by red and orange arrows. The transitions constituting the cycle are colored black and red and the numbers next to them indicate the soft spots switching in the transition. Red numbers indicate soft spots activating in important transitions outside the cycle. (b) The spatial position of the soft spots activated during the cycle. The ellipses enclose the four center particles of each soft spot, where the different colors aim to distinguish between overlapping ellipses. (c) The difference in $\Gamma^+_i$ (black) and $\Gamma^-_i$ (red) values of the multi-periodic soft spots in the three driving cycles.} 
\label{Fig7}
\end{center}
\end{figure*}

In Fig.~\ref{Fig7}(a) we show the graph of a $T=3$ cycle found in a typical $t$-graph. We see that out of the $21$ soft spots participating in the $T=3$ cycle, seven have period-$1$ and the rest are multi-periodic with various switching patterns. 
The mechanisms leading to multi-periodicity of soft spots are similar to the ones found in the $T=2$ cycle considered in the main text.
Some of these soft spots show characteristics that indicate that they may be frozen in part of the cycle. Two examples are soft spots \textit{11} and \textit{16}, for which $|\Gamma^\pm_{11}| \ll \Gamma_{\max}$ and $|\Gamma^\pm_{16}| \ll \Gamma_{\max}$. Note that both appear only in one of the three cycles.

In Fig.~\ref{Fig7}(c), we show the strain thresholds of some soft spots exhibiting threshold oscillations. Here we see similar patterns of threshold oscillations around $\pm\Gamma_{\max}$, \textit{e.g.} $\Gamma^+_{19} > \Gamma_{\max}$ in two out of three cycles, $\Gamma^+_{6}, \Gamma^+_{8} > \Gamma_{\max}$ in one out of three cycles, and $\Gamma^-_{13} < -\Gamma_{\max}$ in two out of three cycles. Soft spot \textit{15} follows a switching pattern that is not applicable in $T=2$ cycles; $\Gamma^+_{15}$ oscillates around $\Gamma_{\max}$ and $\Gamma^-_{15}$ oscillates around $-\Gamma_{\max}$ in different parts of the cycle.
However, due to the larger number of soft spots and more complex dynamics stemming from the period-3 dynamics, we were not able to pin-point a specific example of cascade-scrambling.

Similarly to the $T=2$ cycle shown in Fig.~\ref{Fig1}, in Fig.~\ref{Fig7}(b) the soft spots are localized at the top right corner of the simulation square. However, due to the larger number of soft spots switching, this effect is more pronounced when $T=3$. The larger density of soft spots causes some of them to overlap, which may affect their dynamics. 
We note that some overlalp also exists in Fig.~\ref{Fig1}.
However, due to the small soft spot density in the $T=2$ cycle, any effect that such overlap has on the dynamics can be ignored.

\section{Conditions for threshold oscillations of a $3$-hysteron system}
\label{AppD}
Here we formulate a set of conditions for the occurrence of a $T=2$ cycle in the three-hysteron system discussed in Sec.~\ref{section_4}.
In order for three hysterons to form a $T=2$ cycle, they should form sequential threshold oscillations, \textit{i.e.} hysteron \textit{1} affects \textit{2} that affects \textit{3}, which finally affects \textit{1}, closing the loop. This behavior enforces constraints on the initial values of the strain thresholds $\gamma^{\pm}_i$ of the hysterons.
A sufficient set of requirements for the bare thresholds are that two hysterons will have $\gamma^+ = \Gamma_{\max} \pm \delta$ and the third have $\gamma^- = -\Gamma_{\max} \pm \delta$ (or equivalently two hysterons with $\gamma^- = -\Gamma_{\max} \pm \delta$ and the third with $\gamma^+ = \Gamma_{\max} \pm \delta$), where $\delta$ is a small positive number $\delta/\Gamma_{\max}\ll 1$.
For simplicity we will start the cycle from ${\bf S}=(0,0,0)$ and follow the example given in Fig.~\ref{Fig4}. However, starting from a different initial state will enforce additional constraints based on the switching order of the hysteron in the first driving cycle. 

We will mark the dressed state of the $i$th hysteron after $n$ transition events, \textit{i.e.} simulation steps, as
\begin{equation}
\label{simple-1}
\Gamma^\pm_i(n) = \gamma^\pm_i + \sum_{j \neq i}{s_j(n) \epsilon_j A^{(\pm)}_{i,j}},
\end{equation}
where $s_j(n)$ is the state of hysteron $j$ after $n$ transitions.
For the $\nearrow$ segment of the first driving cycle, and without a loss of generality, the nearest threshold (at $n=0$) is:
\begin{equation}
\label{simple-2}
\min{\left\{ \gamma^+_1, \gamma^+_2, \gamma^+_3, \Gamma_{\max}\right\}} = \gamma^+_3  \equiv \Gamma^+_3(0),
\end{equation} 
which means that soft spot \textit{3} switches first. In order to have a $T=2$ cycle, \textit{2} must not switch. Therefore, for $n=1$, the threshold is:
\begin{equation}
\label{simple-3}
\min{\left\{ \Gamma^+_1(1), \Gamma^+_2(1), \Gamma_{\max}\right\}} = \Gamma^+_1(1),
\end{equation} 
as shown in the example Fig~\ref{Fig4}b.
This means that hysteron \textit{1} switches second. At $n=2$, we have:
\begin{equation}
\label{simple-4}
\min{\left\{ \Gamma^+_2(2), \Gamma_{\max}\right\}} = \Gamma_{\max}.
\end{equation}
and therefore \textit{2} does not switch. 
To clarify, the strain threshold of hysteron \textit{2} can be lower than $\Gamma_{\max}$ at some point of the $\nearrow$ segment, however, it must be above the strain thresholds of hysterons \textit{1} and \textit{3}, and be pushed above $\Gamma_{\max}$ in the process.

We next reach the $\searrow$ segment of the first cycle and for $n=2$, we get that the first hysteron to switch is:
\begin{equation}
\label{simple-5}
\max{\left\{ \Gamma^-_1(2), \Gamma^-_3(2), -\Gamma_{\max}\right\}} = \Gamma^-_3(2),
\end{equation} 
and since \textit{2} is in the state $s_2=0$ it cannot switch in the $\searrow$ segment of decreasing strain, we must have that:
\begin{equation}
\label{simple-6}
	\Gamma^-_1(3) = \gamma^-_1 < -\Gamma_{\max}.
\end{equation} 
in order for the cycle to have $T=2$.

In the second driving cycle there is a change to the switching order due to hysteron \textit{1}, where now \textit{2} must switch first since if \textit{3} switches first, the system returns to its previous state, therefore,
\begin{equation}
\label{simple-7}
\min{\left\{ \Gamma^+_2(3), \Gamma^+_3(3), \Gamma_{\max}\right\}} = \Gamma^+_2(3).
\end{equation}
Two options are available to close the threshold oscillation loop, depending on whether $\Gamma^+_3(4)$ is greater or smaller than $\Gamma_{\max}$. In this example 
\begin{equation}
\label{simple-8}
\min{\left\{ \Gamma^+_3(4), \Gamma_{\max}\right\}} = \Gamma_{\max}.
\end{equation}
The transitions in $\searrow$ segment of the second cycle are then forced:
\begin{equation}
\label{simple-9}
\max{\left\{ \Gamma^-_1(4), \Gamma^-_2(4), -\Gamma_{\max}\right\}} = \Gamma^-_1(4),
\end{equation}
and finally
\begin{equation}
\label{simple-10}
\Gamma^-_2(5) = \gamma^-_2 >  -\Gamma_{\max},
\end{equation}
returns to the initial state.

In this example we see that for the case of a three-hysteron system, once the initial state of the system is set, the relations between the strain thresholds of the hysterons are also determined, and in order to achieve multi-periodicity, the interactions between them must be carefully chosen.

\section{Destabilizing interactions and multi-periodicity}
\label{AppE}
In many systems, such as the random field Ising model (RFIM) charge density waves (CDW) and the original Preisach model, the interactions are strictly destabilizing. This means that a change of state of hysteron $j$, $s_j=0\rightarrow 1$ or $s_j=1\rightarrow 0$ can only cause the activation threshold of hysteron $i$, $|\Gamma^{\pm}_i|$ to remain the same, or become smaller with respect to $\Gamma_{\max}$. 
In such systems, the no-passing theorem guarantees that they can never have periodic cycles of $T>1$ \cite{Middleton1992,Sethna1993}. Here we show directly that such interactions preclude periodic cycles with period larger than one, but that allowing the interactions to be stabilizing, allows such cycles.\\

{\it Proof that non-monotonic interactions are necessary to have period-2 in the thee hysteron system}:

In the three-hysteron example, because $T=2$, both {\bf s}$_a=(1,1,0)$ and {\bf s}$_b=(1,0,1)$ are stable at $\Gamma=\Gamma_{\max}$.
Now, assume that the interactions are always destabilizing.
In the case of iPreisach this means that:
\begin{equation}
\epsilon_iA_{ij}^{(+)} <0
\end{equation}
for every $i$ and $j$. 
For configurations $a$ and $b$, we must have that:
\begin{equation}
\Gamma^{a,+}_2 < \Gamma_{max} < \Gamma^{b,+}_{2}
\label{condition}
\end{equation}
since s$_2^a=1$ and s$_2^b=0$ at $\gamma=\Gamma_{max}$.
However, due to interactions, the values of  $\Gamma^{a,+}_2,\Gamma^{b,+}_2$ are also affected by the states of hysterons $1$ and $3$:
\begin{equation}
\Gamma^{a,+}_2 = \gamma^+_2 + \epsilon_1A_{12}^{(+)}
\end{equation}
\begin{equation}
\Gamma^{b,+}_2 = \gamma^+_2 + \epsilon_1A_{12}^{(+)} + \epsilon_3A_{32}^{(+)}
\end{equation}
subtracting the equations, we get that:
\begin{equation}
\Gamma^{b,+}_2 - \Gamma^{a,+}_2 =  \epsilon_3A_{32}^{(+)}< 0
\end{equation}
and thus:
\begin{equation}
\Gamma^{a,+}_2 > \Gamma^{b,+}_2
\end{equation}
in contradiction to condition \ref{condition}. However, if $\epsilon_iA_{ij}^{(+)}$ are allowed to be both positive and negative, condition \ref{condition} can still be fulfilled and periods larger than one are possible, as we saw in the example. This can be generalized to systems of any number of hysterons.

\end{document}